\documentclass[a4paper,11pt]{article}
\pdfoutput=1 

\usepackage{jheppub} 
\usepackage{amsmath}

\usepackage{caption}
\usepackage{subcaption}
\usepackage{ dsfont }
\usepackage{nicematrix}
\usepackage{ bbold }
\usepackage{hyperref}

\usepackage[T1]{fontenc} 
\newcommand{\comment}[1]{}
\usepackage{booktabs}
\usepackage{siunitx}

\newcommand{\eqsplit}[2][]{%
    \begin{equation}\label{#1}
    \begin{split}
    #2
    \end{split}
    \end{equation}%
}

\newcommand{\doubletraceone}{\vcenter{\hbox{$\mathord{\includegraphics[width=7.2ex]{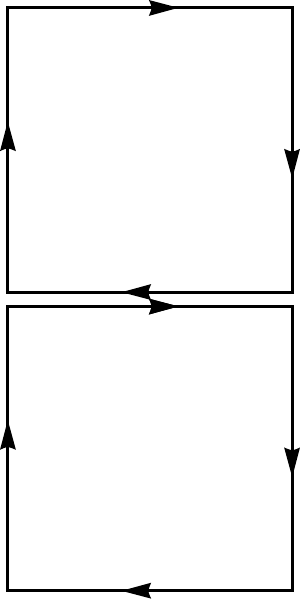}}$}}}
\newcommand{\doubletracetwo}{\vcenter{\hbox{$\mathord{\includegraphics[width=7.2ex]{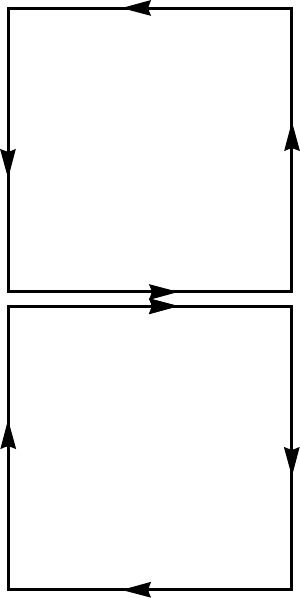}}$}}}
\newcommand{\doubletracethree}{\vcenter{\hbox{$\mathord{\includegraphics[width=7.2ex]{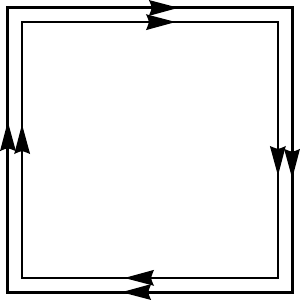}}$}}}
\newcommand{\doubletracefour}{\vcenter{\hbox{$\mathord{\includegraphics[width=7.2ex]{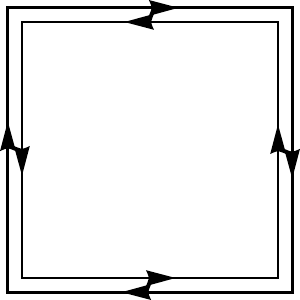}}$}}}

\newcommand{\wone}{\vcenter{\hbox{$\mathord{\includegraphics[width=4.8ex]{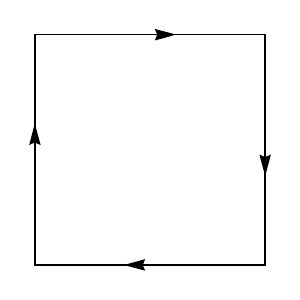}}$}}}
\newcommand{\wtwo}{\vcenter{\hbox{$\mathord{\includegraphics[width=7.2ex]{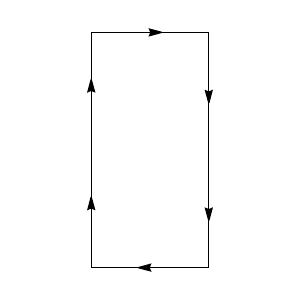}}$}}}
\newcommand{\wthree}{\vcenter{\hbox{$\mathord{\includegraphics[width=4.8ex]{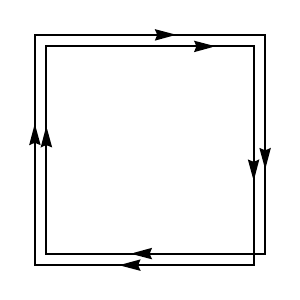}}$}}}
\newcommand{\wfour}{\vcenter{\hbox{$\mathord{\includegraphics[width=7.2ex]{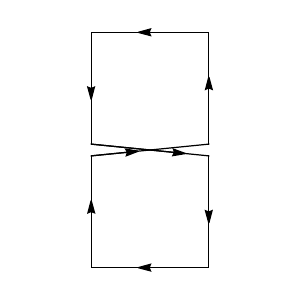}}$}}}
\newcommand{\wfive}{\vcenter{\hbox{$\mathord{\includegraphics[width=7.2ex]{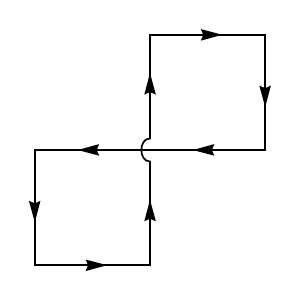}}$}}}
\newcommand{\wsix}{\vcenter{\hbox{$\mathord{\includegraphics[width=7.2ex]{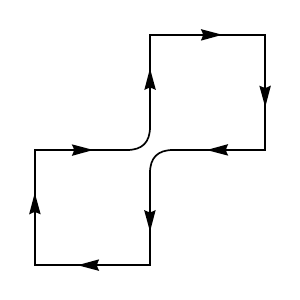}}$}}}
\newcommand{\lineone}{\vcenter{\hbox{$\mathord{\includegraphics[width=7.2ex]{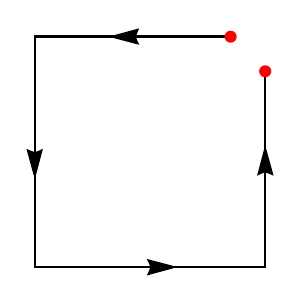}}$}}}
\newcommand{\linetwo}{\vcenter{\hbox{$\mathord{\includegraphics[width=7.2ex]{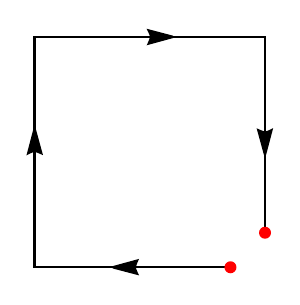}}$}}}
\newcommand{\linethree}{\vcenter{\hbox{$\mathord{\includegraphics[width=7.2ex]{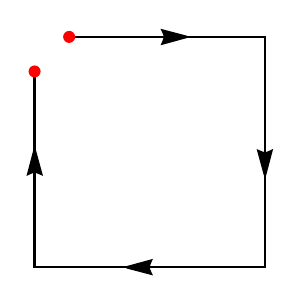}}$}}}
\newcommand{\linefour}{\vcenter{\hbox{$\mathord{\includegraphics[width=7.2ex]{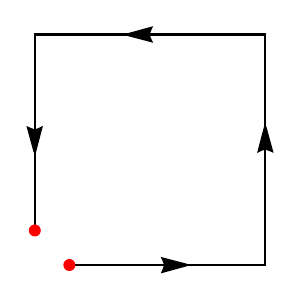}}$}}}
\newcommand{\linefive}{\vcenter{\hbox{$\mathord{\includegraphics[width=7.2ex]{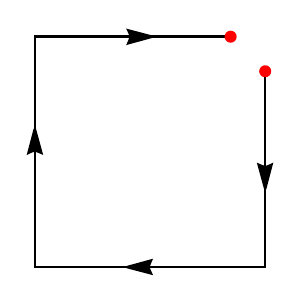}}$}}}
\newcommand{\linesix}{\vcenter{\hbox{$\mathord{\includegraphics[width=7.2ex]{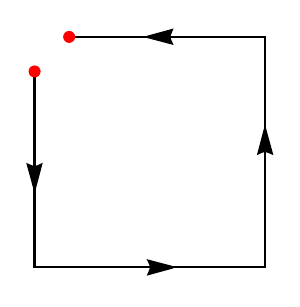}}$}}}
\newcommand{\lineseven}{\vcenter{\hbox{$\mathord{\includegraphics[width=7.2ex]{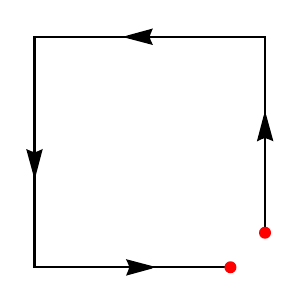}}$}}}
\newcommand{\lineeight}{\vcenter{\hbox{$\mathord{\includegraphics[width=7.2ex]{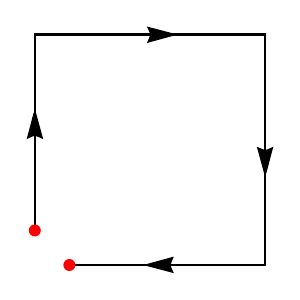}}$}}}

\newcommand{\redundantone}{\vcenter{\hbox{$\mathord{\includegraphics[width=14.4ex]{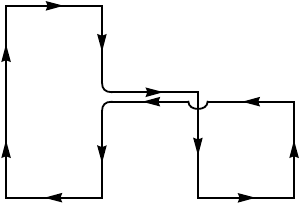}}$}}}
\newcommand{\redundanttwo}{\vcenter{\hbox{$\mathord{\includegraphics[width=14.4ex]{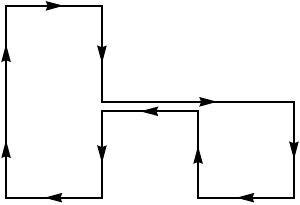}}$}}}
\newcommand{\redundantthree}{\vcenter{\hbox{$\mathord{\includegraphics[width=14.4ex]{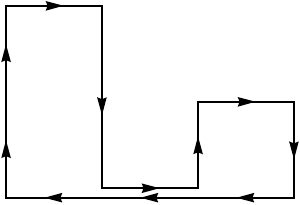}}$}}}
\newcommand{\redundantfour}{\vcenter{\hbox{$\mathord{\includegraphics[width=14.4ex]{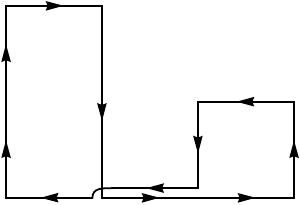}}$}}}

\newcommand{\doubleone}{\vcenter{\hbox{$\mathord{\includegraphics[width=14.4ex]{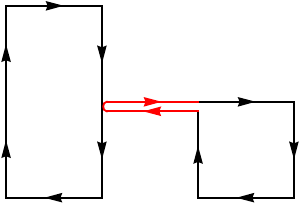}}$}}}
\newcommand{\doubletwo}{\vcenter{\hbox{$\mathord{\includegraphics[width=14.4ex]{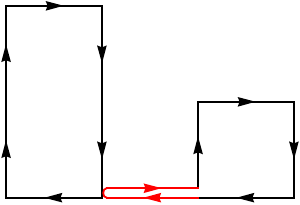}}$}}}

\newcommand{\doublezero}{\vcenter{\hbox{$\mathord{\includegraphics[width=14.4ex]{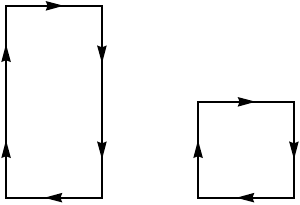}}$}}}

\newcommand{\levelone}{\vcenter{\hbox{$\mathord{\includegraphics[width=10ex]{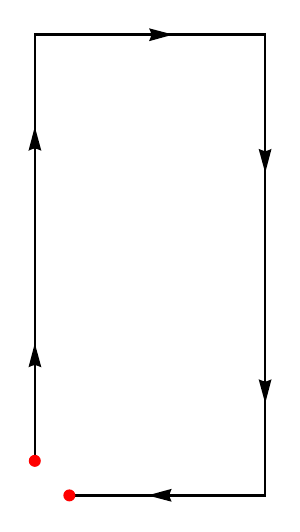}}$}}}
\newcommand{\leveltwo}{\vcenter{\hbox{$\mathord{\includegraphics[width=10ex]{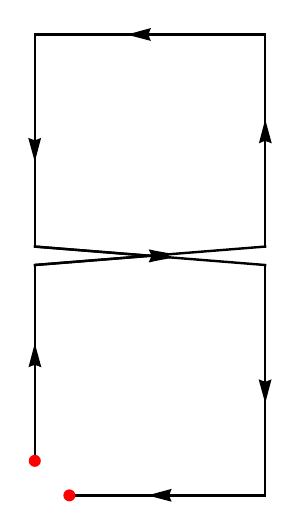}}$}}}
\newcommand{\levelthree}{\vcenter{\hbox{$\mathord{\includegraphics[width=18ex]{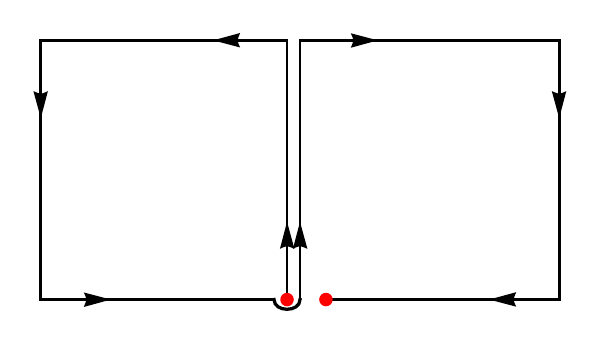}}$}}}
\newcommand{\levelfour}{\vcenter{\hbox{$\mathord{\includegraphics[width=18ex]{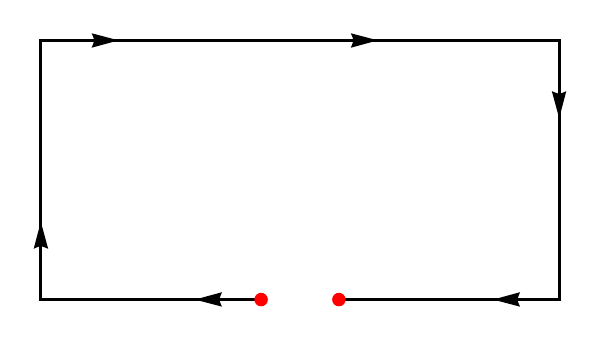}}$}}}
\newcommand{\levelfive}{\vcenter{\hbox{$\mathord{\includegraphics[width=14.4ex]{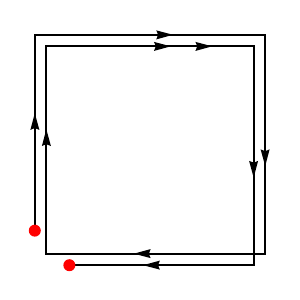}}$}}}

\newcommand{\p}{\partial}
\newcommand{\e}{\mathrm{e}}
\newcommand{\Det}{{\rm Det }}
\newcommand{\tr}{{\rm tr }}
\newcommand{\Tr}{{\rm Tr }}

\title{\boldmath Bootstrap for Finite N Lattice Yang-Mills Theory}

\author[a]{Vladimir Kazakov}
\author{{\it and}}
\author[b,a]{Zechuan Zheng}

\affiliation[a]{Laboratoire de Physique de l'\'Ecole normale sup\'erieure,\\
ENS, Universit\'e PSL, CNRS, Sorbonne Universit\'e,\\
Universit\'e de Paris, 24 rue Lhomond, 75005 Paris, France
}
\affiliation[b]{Perimeter Institute for Theoretical Physics,
Waterloo, ON N2L 2Y5, Canada}

\emailAdd{Vladimir.Kazakov@ens.fr}
\emailAdd{zechuan.zheng.phy@gmail.com}

\abstract{
We introduce a comprehensive framework for analyzing finite \(N\) lattice Yang-Mills theory and finite \(N\) matrix models. Utilizing this framework, we examine the bootstrap approach to SU(2) Lattice Yang-Mills Theory in 2,3 and 4 dimensions.  The SU(2) Makeenko-Migdal loop equations on the lattice are linear and closed exclusively on single-trace Wilson loops. This inherent linearity significantly improves the efficiency of the bootstrap approach by leveraging the problem’s convexity, permitting the inclusion of Wilson loops up to length 24. The exact upper and lower margins for the free energy per plaquette, derived from our bootstrap method, demonstrate good agreement with Monte Carlo data, achieving precision within \(0.1\%\) for the physically relevant range of couplings in both three and four dimensions. Additionally, our bootstrap data provides estimates of the string tension, in qualitative agreement with existing Monte Carlo computations.
}

\begin{document} 
\maketitle
\flushbottom

\section{Introduction}

A great deal of our knowledge of the non-perturbative properties of the gauge theories in physical three and four space-time dimensions is based on the  Lattice Monte Carlo calculations of the lattice Yang-Mills theories\cite{Creutz:1980zw, Creutz:1980wj}. The current state of the art in this area, thanks to the intensive use of supercomputers and modern algorithms, allows to achieve remarkable precision for the computation of spectra of glueballs and mesons (when the quark fields are included), with the size of lattices now attaining hundreds of lattice units~\cite{Athenodorou:2021qvs}. On the other hand, relying solely on this method is not entirely satisfactory from both theoretical and practical perspectives.  The search for other universal non-perturbative approaches did not lose its relevance. 

Recent advancements in the field\cite{Anderson:2018xuq, Lin:2020mme, Han:2020bkb, Kazakov:2021lel, Kazakov:2022xuh} have given rise to an innovative bootstrap method for the analysis of matrix models and lattice theories. This approach is unique in that it not only adheres to standard assumptions, like unitarity and global symmetries, but also incorporates constraints among physical observables that are dictated by the equations of motion. This methodology promises a more rigorous framework for the exploration of matrix model dynamics. 
The broad utility of this novel bootstrap approach has been rapidly demonstrated across various domains, including lattice field theories \cite{Anderson:2016rcw, Anderson:2018xuq, Cho:2022lcj, Kazakov:2022xuh}, matrix models \cite{Han:2020bkb, Jevicki:1982jj, Jevicki:1983wu, Koch:2021yeb, Lin:2020mme, Lin:2023owt, Mathaba:2023non}, quantum systems \cite{Aikawa:2021eai, Aikawa:2021qbl, Bai:2022yfv, Berenstein:2021dyf, Berenstein:2021loy, Berenstein:2022unr, Berenstein:2022ygg, Berenstein:2023ppj, Bhattacharya:2021btd, Blacker:2022szo, Ding:2023gxu, Du:2021hfw, Eisert:2023hcx, Fawzi:2023ajw, hanQuantumManybodyBootstrap2020, Hastings:2021ygw, Hastings:2022xzx, Hessam:2021byc, Hu:2022keu, Khan:2022uyz, Kull:2022wof, Li:2022prn, Li:2023nip, Guo:2023gfi, Morita:2022zuy, Nakayama:2022ahr, Nancarrow:2022wdr, Tavakoli:2023cdt, Tchoumakov:2021mnh, Fan:2023bld, Fan:2023tlh, John:2023him, Li:2023ewe, Zeng:2023jek, Fawzi:2023fpg, Li:2024rod}, and even classical dynamical systems \cite{goluskinBoundingAveragesRigorously2018, goluskinBoundingExtremaGlobal2020, tobascoOptimalBoundsExtremal2018, Cho:2023xxx}. This extensive range of applications highlights the versatility and robustness of the method, underlining its significance in advancing theoretical frameworks across these fields.

In our previous work~\cite{Kazakov:2022xuh}, we employed the bootstrap approach to study the $SU(N)$ lattice Yang-Mills theories at $N=\infty$ in various dimensions, using the lattice Makeenko-Migdal equation (MME) for the Wilson loop~\cite{Makeenko:1979pb}. Initial developments in this area are due to~\cite{Anderson:2016rcw}. In this planar limit, the loop equations are inherently non-linear, leading in the context of bootstrap to a non-convex optimization problem. To address this complexity, we implemented a relaxation procedure in~\cite{Kazakov:2021lel}.

In this paper, we generalize the bootstrap procedure for the matrix models and lattice gauge theories in the planar limit to the finite \(N\) situation. First, we notice that, while the multi-trace averages are important to formulate a closed system of MMEs, we can explore a general property of the matrix-valued theories that the so-called trace identities help to reduce the number of traces in the multi-trace products of matrix fields. That is, the product of $k>N$ traces (or $k>N-1$, depending on the symmetries) of arbitrary products of matrices can be represented as the sum of products of smaller numbers of traces. Higher trace variables are described through trace identities specific to the corresponding matrix ensemble. 

In this paper, we concentrate on systems with variables SU(2) and SU(3), where the loop equations appear to be linear in the space of single-trace and double-trace averages, respectively. For the  SU(3)  case, we discuss the general framework of linear double-trace loop equations, including the lattice gauge theory case, and solve via bootstrap the one-matrix model for illustrative purposes. The rest of our paper is concentrated on the study and bootstrap solution of the loop equation for SU(2) lattice Yang-Mills theory. We extensively use the fact that, in any dimension, this equation closes exclusively on the single trace Wilson loops and is fully linear~\footnote{as noticed already in~\cite{Gambini:1990dt}}. This linearity obviates the need for the relaxation procedures typically required in the large N limit. With the aid of the hierarchy of Wilson loops discussed in this article, along with the specialized program we developed, we have maximized the capabilities of the bootstrap procedure. The linearity of MMEs helped us to exploit the full power of the utilized hardware and of the advanced optimization software, allowing us to bootstrap the Wilson loops up to a maximum length of 24 lattice units.

We believe that the linear SU(2) loop equations in the single-trace loop space which we derive in detail in this paper, both on the lattice and in the continuum,  can provide important insights into the nature and properties of the Yang-Mills theory. We will mostly concentrate here on the practical aspects and advantages of such linearity for the bootstrap procedure.

\subsection{Main results}

\begin{figure}[htp]

\subfloat[]{%
\centering
  \includegraphics[clip,width=.7\columnwidth]{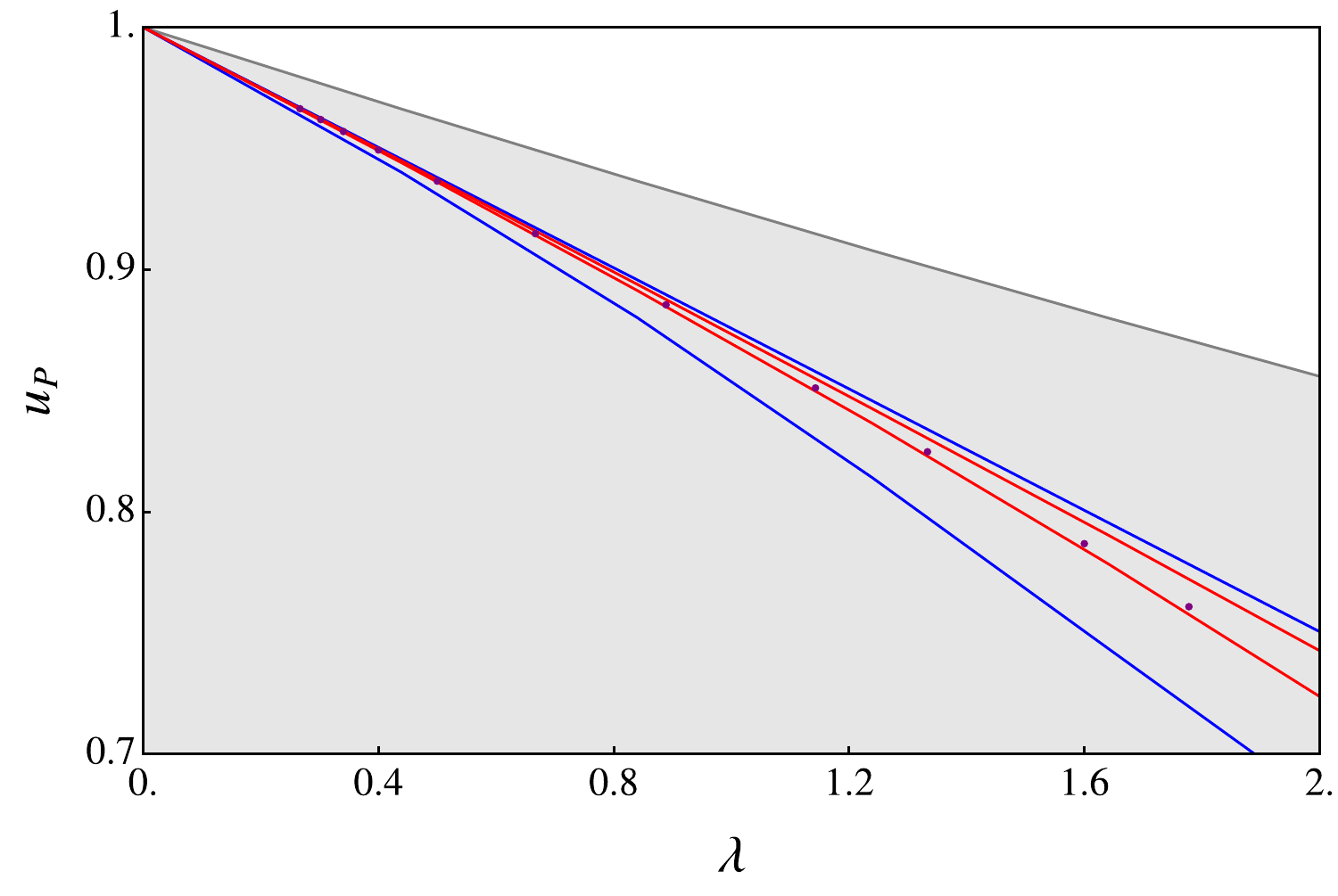}
}

\subfloat[]{%
\centering
  \includegraphics[clip,width=.7\columnwidth]{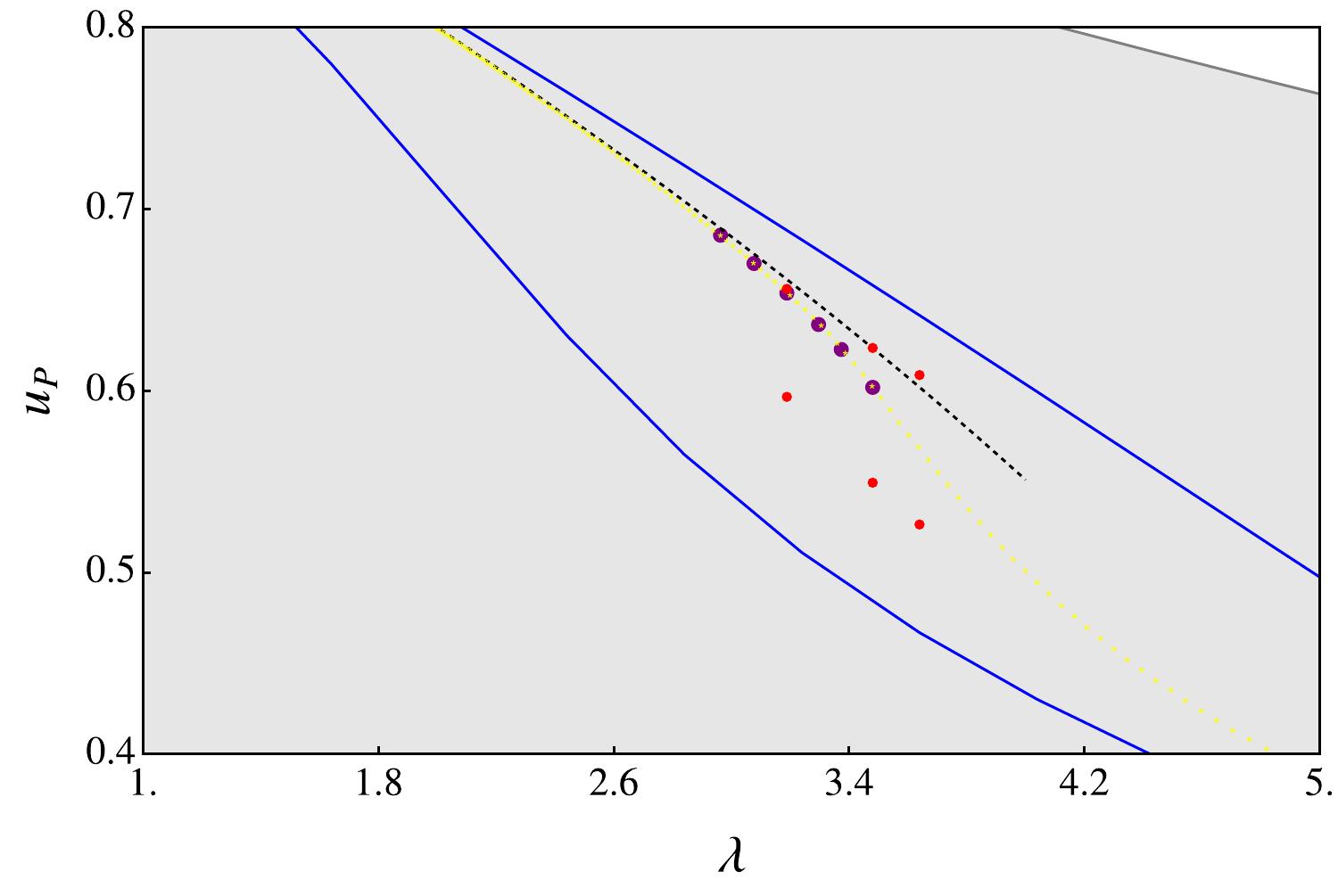}\label{fig: intro4}
}
\caption{The results of the bootstrap calculations in three and four dimensions for the plaquette average \( u_{P} = \tr U_{\Box} \) as a function of the coupling \( \lambda = \frac{2N^2}{\beta} \) are displayed graphically. Blue lines represent interpolations of the upper and lower margins for the maximum length of loops \( L_{\text{max}} \leq 16 \) used in the bootstrap procedure for loop equations. Red lines and dots illustrate the same for \( L_{\text{max}} \leq 24 \). The gray area indicates the allowed values of \( u_{P} \) for \( L_{\text{max}} \leq 8 \). For additional details regarding these plots, refer to Section~\ref{sec: higherdim}. The dashed lines in the figure represent the strong and weak coupling expansions, extracted from \cite{Denbleyker:2008ss}. Additionally, the results from Monte Carlo simulations are depicted using purple dots \cite{Athenodorou:2016ebg,Athenodorou:2021qvs} and yellow dots \cite{Denbleyker:2008ss}, each set representing data from different sources. For additional details regarding these plots, refer to Section~\ref{sec: higherdim}.}
\label{fig: intro}
\end{figure}

This paper proposes a general framework for bootstrapping finite \(N\) lattice Yang-Mills theory and finite \(N\) matrix models, highlighting the crucial role of multi-trace variable truncation. Specifically, the number of independent multi-trace variables is bounded, with higher-number trace variables expressed as linear combinations of a subset of multi-trace variables. These truncations, allowing reduction of any multi-trace SU($N$) variables to a linear combination of only $N-1, N-2,\dots,1$-trace variables,    are demonstrated here through the following trace identities for SU(2) and SU(3):
\begin{equation}\label{eq: SU(2)trace}
    \tr U \tr V = \tr UV + \tr U V^\dagger, \quad \forall U, V \in SU(2)
\end{equation}
\begin{equation}\label{eq: SU(3)trace}
\begin{split}
&\tr U \tr V \tr W = -\tr W V U - \tr U V W + \tr U \tr V W + \tr U  V \tr W + \tr U W \tr V \\
&+\tr U^\dagger V \tr U^\dagger W - \tr U^\dagger V U^\dagger W, \qquad\qquad \forall U, V, W \in SU(3)
\end{split}
\end{equation}

\begin{figure}[t]
\begin{center}
\includegraphics[scale=0.4]{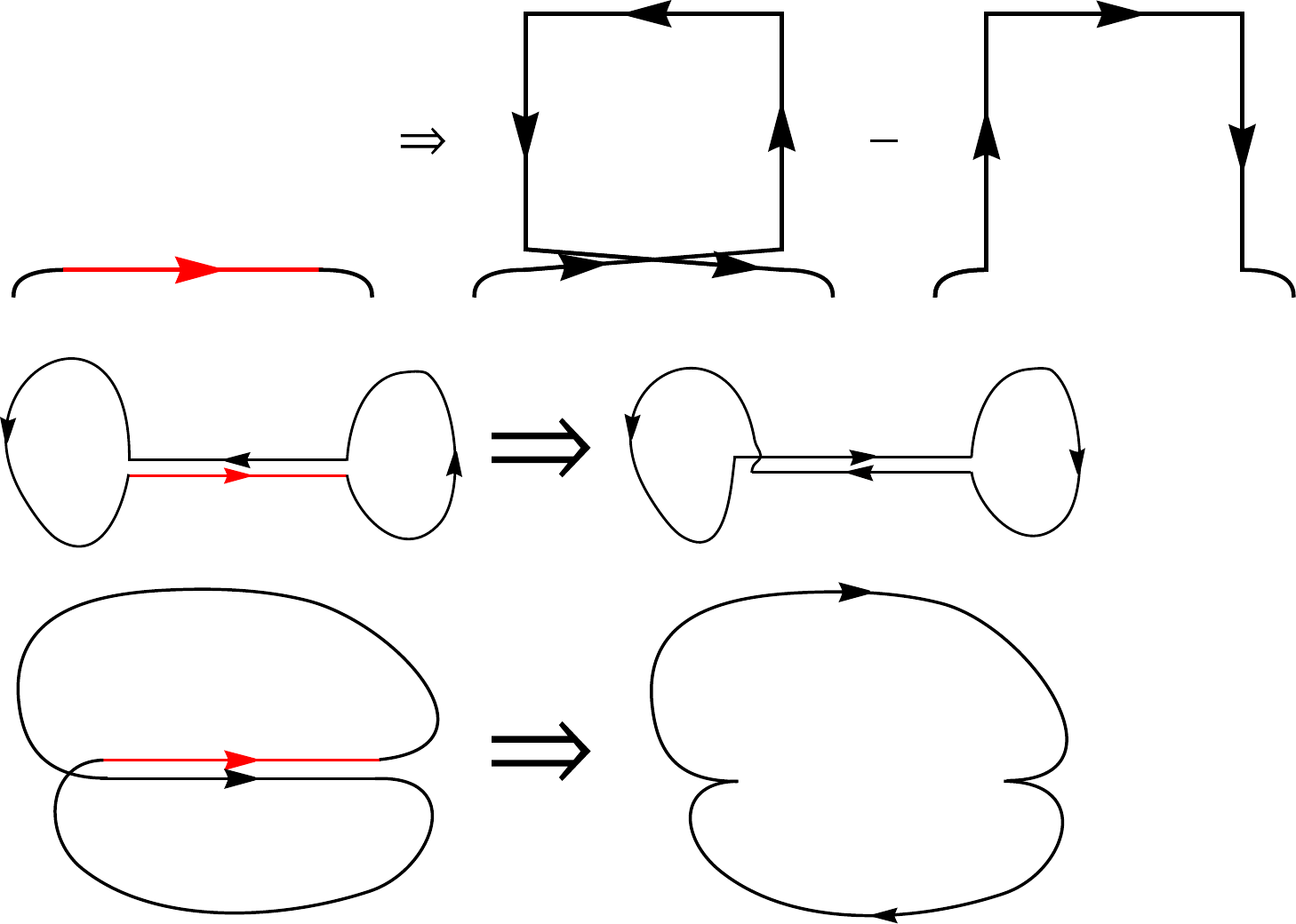}
\end{center}
\caption{Illustration of the different terms in the loop equation \eqref{eq: su2loopeqcompact}.}  
\label{fig: loopvar2d}
\end{figure}

We then explored the dynamics of multi-trace Wilson loops, captured by multi-trace loop equations, both on the lattice and in the continuum. For SU(2), the trace identities simplify the representation of every double-trace Wilson loop to a sum of single-trace Wilson loops defined as an averaged trace of the product of group variables $U_{l_{\nu}}$ on the links $l_{\nu}$ along the lattice contour $C$\footnote{Here and in the whole article, we use the notation that $\Tr =\frac{1}{N}\tr$, so the normalization of the identity equation has $\Tr \mathbb{1}=1$.}:
\begin{align}
    W(C)=\langle \Tr \prod_{l_{\nu}\in C} U_{l_{\nu}} \rangle\,.
\end{align}
The specific form of the SU(2) Makeenko-Migdal loop equations for the Wilson lattice action~\eqref{eq: actionYM}   is compactly represented as:
\begin{align}\label{eq: su2loopeqcompact}
\frac{1}{\lambda}\sum_{\mu\neq \pm\nu} \Big[ W(\sigma^{l}_{\nu\mu}\circ C_{l_{\nu}})-W(\sigma^{l}_{\mu\nu}\circ C_{_{l_{\nu}}})\Big]+\frac{1}{2}\sum_{l'_{\nu}\sim l_{\nu}\in C} (-1)^{\frac{1+\nu\cdot\nu'}{2}}\left(W(C^{(1)}_{x,\nu}\circ \overline{C^{(2)}_{y,\nu'}})+\frac{1}{2}W(C)\right)=0   
\end{align}
where $C\circ C'$ denotes the composition of two contours (joint at a given common link), $\sigma^{l}_{\nu\mu}$ is a plaquette which includes the link $l$ and has the orientation $\nu,\mu$, where any Greek index takes the values $\pm 1,\pm 2\,\dots,\pm d$ (sign is related to the orientation of the link w.r.t gauge variable), $\overline{C}$ means the contour $C$ with the flipped direction  (conjugation of the corresponding Wilson path). The notation \(\sum_{l'_{\nu} \sim l_{\nu}}\) denotes the summation over links in the contour \(C\) that coincide with \(l_{\nu}\) on the lattice. Importantly, this summation explicitly includes \(l_{\nu}\) itself. The variations and specific instances of this summing process are visually depicted in Figure~\ref{fig: loopvar2d}, while a more detailed and explicit mathematical formulation can be found in Section~\ref{sec: loopeq}.

We also derive the finite \(N\) Makeenko-Migdal equation in the continuum, focusing here on the SU(2) case, where the continuous loop equation is~\cite{Gambini:1990dt} \footnote{Derivation of such equations on the lattice and in continuum, and in particular of the SU(3) loop equation,  is deferred to Section~\ref{sec: loopeq}}: 
\begin{align}
\frac{\p}{\p x_\mu}\frac{\delta}{\delta\sigma_{\mu\nu}(x)}W(C_x)=\frac{\lambda}{2}\oint_ {C}dy_{\nu}\,\delta^{(d)}(x-y)\left[\,\,\frac{1}{2}W(C_x)+W\left(C^{(1)}_{xy}\circ \overline{C^{(2)}_{yx}}\right)\right]
\end{align}

This situation is simpler than that of thepreviously studied  't Hooft limit, where double-trace Wilson loops are products of single-trace Wilson loops due to large \(N\) factorization. There is no need for relaxation in the bootstrap scheme, which is typically required due to the non-convex nature of the large \(N\) loop equations. The inherent linearity of the SU(2) loop equations renders the entire bootstrap procedure convex from the very beginning.

The paper concludes with a detailed examination of SU(2) lattice Yang-Mills theory across two, three, and four dimensions. A major technical achievement of this work is the significant extension of the maximum length of Wilson loops involved in our bootstrap procedure — up to 24 lattice units for the 2D, 3D and 4D cases. The hierarchical system, which categorizes Wilson lines by level and selects operators based on this categorization, has markedly increased the precision of computations for Wilson averages of small loops. Notably, the margins for the plaquette average in three and four dimensions within the physically relevant coupling range achieve a precision of the order of 0.1\%. These findings are depicted in Figure~\ref{fig: intro}, showcasing results for three and four dimensions, respectively. Additionally, using interpolation of the Creutz ratio, we extracted string tension data. Although these results do not match the precision of Monte Carlo simulations due to the limited loop size (maximum \(2\times3\)), they are qualitatively sound.

\textbf{Note added}: While we were finalizing this article, we became aware of the work \cite{Li:2024wrd}, which focuses on the Abelian gauge theory and parallels some of our discussions on the one-matrix model.

\section{Loop equations}\label{sec: loopeq}

Our bootstrap approach is aimed at the expectation values of multi-trace quantities within the given matrix ensemble. Our method differs from the schemes involving the 't Hooft limit, where, due to the large \(N\) factorization of multi-trace averages, the loop equations close on the single-trace space\footnote{For more complex physical quantities, such as glueball correlation functions, multi-trace averages are necessary.}~\cite{Anderson:2018xuq, Kazakov:2022xuh}. 

Although at finite \(N\) we cannot factorize the the multi-trace averages into products of averages of single traces, we can make use of trace identities to reduce such averages to a limited number of traces inside the averages.  Several of the trace identities discussed in this section were originally derived in \cite{Migdal:1983qrz, Gambini:1986ew}. These identities effectively limit the space of multi-trace averages to those containing no more than \(N\) traces, with the specific reduction contingent upon the random matrix ensemble being studied\footnote{It will be beneficial to further clarify the relationship between the trace identities presented in our study and those associated with finite \(N\) discussed in \cite{Dempsey:2022uie, Dempsey:2023fvm}. }. 

The trace identities strongly influence the kinematics of loop variables at finite $N$ and their structure  depends on the specific value of \(N\) in a non-trivial way. As the first step of our approach, we derive the multi-trace loop equations, which concern the dynamics of the loop variables. We will focus on two models: initially, the SU(3) one-matrix model, which will serve as an illustrative example, followed by the SU(2) lattice Yang-Mills theory employing the Wilson action. 

\subsection{Finite N trace identities}\label{sec: trace}

The trace identities we consider here are quite general. We begin with a discussion of the case of general matrices in \(GL(N)\). An useful observation follows:
\begin{equation}
\epsilon^{a_1a_2\ldots a_{N+1}}\epsilon^{b_1b_2\ldots b_{N+1}} M_1^{a_1 b_1}\ldots M_{N+1}^{a_{N+1} b_{N+1}}=0,\quad \forall\, M_1,\ldots,M_{N+1} \in GL(N)
\end{equation}
Additionally, we can express the product of the \(\epsilon\) tensors in terms of the Kronecker delta symbols\cite{enwiki:1243909782}:
\begin{equation}\label{glnepsilon}
    \epsilon^{a_1a_2\ldots a_{N+1}}\epsilon^{b_1b_2\ldots b_{N+1}}=\delta_{a_1b_1}\delta_{a_2b_2}\ldots\delta_{a_{N+1}b_{N+1}}+\sum_{P\neq \mathbf{id}} (-1)^P \delta_{a_1P(b_1)}\delta_{a_2P(b_2)}\ldots\delta_{a_{N+1}P(b_{N+1})}
\end{equation}
The sum runs over all permutations of the \(b\) indices, excluding the identity permutation. Combining these formulas, the first term in Equation \ref{glnepsilon} leads to a \(N+1\) trace variable, while the remainder includes terms with \(N\) traces or fewer:
\begin{equation}
    \tr M_1 \tr M_2 \ldots\tr M_{N+1}=-\sum_{P\neq \mathbf{id}} (-1)^P \delta_{a_1P(b_1)}\delta_{a_2P(b_2)}\ldots\delta_{a_{N+1}P(b_{N+1})}M_1^{a_1 b_1}\ldots M_{N+1}^{a_{N+1} b_{N+1}}
\end{equation}
As an example, for GL(2), we get:
\begin{small}
\begin{equation}\label{eq: GL(N)}
    \tr M_1 \tr M_2 \tr M_3= - \tr M_3 M_2 M_1 -\tr M_1 M_2 M_3+ \tr M_1 \tr M_2  M_3 +\tr M_1  M_2 \tr M_3 +\tr M_1 M_3 \tr M_2
\end{equation}
\end{small}
This indicates that in the case of GL(2), the trace variables generally truncate at double traces.

More interestingly, for the subgroups of GL(N), there is evidence suggesting that the truncation of multi-trace variables can be further refined. For SU(N) and SO(N), the unimodularity condition further applies:
\begin{equation}
    1=\Det U=\Det U^\dagger=\frac{1}{N!} \epsilon^{a_1a_2\ldots a_{N}}\epsilon^{b_1b_2\ldots b_{N}} U^{\dagger a_1 b_1}\ldots U^{\dagger a_{N} b_{N}}
\end{equation}
We multiply  both sides of the previous equation by the following term
\begin{equation}
    \frac{1}{N!} \epsilon^{c_1c_2\ldots c_{N}}\epsilon^{d_1d_2\ldots d_{N}} U^{ c_1 d_1}V_{1}^{ c_2 d_2}\ldots V_{N-1}^{ c_{N} d_{N}}
\end{equation}
and finally, we obtain:
\begin{equation}\label{eq: SU(N)trace}
    \begin{split}
        & \epsilon^{c_1c_2\ldots c_{N}}\epsilon^{d_1d_2\ldots d_{N}} U^{ c_1 d_1}V_{1}^{ c_2 d_2}\ldots V_{N-1}^{ c_{N} d_{N}}\\
    &=\frac{1}{N!} (\epsilon^{a_1a_2\ldots a_{N}}\epsilon^{c_1c_2\ldots c_{N}})(\epsilon^{b_1b_2\ldots b_{N}}\epsilon^{d_1d_2\ldots d_{N}}) U^{ c_1 d_1}V_{1}^{ c_2 d_2}\ldots V_{N-1}^{ c_{N} d_{N}}  U^{\dagger a_1 b_1}\ldots U^{\dagger a_{N} b_{N}}
    \end{split}
\end{equation}
For similar reasons as outlined in \eqref{glnepsilon}, the left-hand side of \eqref{eq: SU(N)trace} contains only one \(N\)-trace variable. Upon contracting the \(\epsilon\) tensors, \(\epsilon^{a_1a_2\ldots a_{N}}\epsilon^{c_1c_2\ldots c_{N}}\) and \(\epsilon^{b_1b_2\ldots b_{N}}\epsilon^{d_1d_2\ldots d_{N}}\), the right-hand side contains at most \((N-1)\)-trace variables. This reduction occurs because \(U\) and \(U^\dagger\) always contract to yield a trivial trace, \(\tr \mathbb{1}\). As examples, for SU(2) and SU(3), we obtain the identities \eqref{eq: SU(2)trace} and \eqref{eq: SU(3)trace} which will be extensively used in this paper.

Therefore, unlike the general case for GL(N) where the trace variables truncate at \(N\) traces, for SU(N) and SO(N), the trace variables truncate at \(N-1\) traces.

\subsection{Loop equations for the one-matrix model}

For illustration purposes, let us first consider the "single plaquette" toy model for the SU(3) matrix ensemble\footnote{We gather a concise summary of the exact solution of this model for SU(N)/U(N) in Appendix~\ref{app: exact}, and provide a general treatment of the SU(N)/U(N) multi-trace loop equations in Appendix~\ref{app: generalloop}.
}:
\begin{equation}\label{eq: onematrixsun}
    Z=\int  \mathcal{D} U\, \e^{\frac{N}{\lambda}(\tr U^\dagger + \tr U)}
\end{equation}
This model represents the simplest non-trivial unitary matrix model, characterized by a real action. Despite its apparent simplicity, the model holds practical physical significance as it describes the dynamics of a single plaquette Wilson loop in a two-dimensional lattice Yang-Mills theory\cite{Gross:1980he, Wadia:2012fr}.

In light of the trace relation \eqref{eq: SU(3)trace} derived in the last section, the space of loop variables truncates at double-trace, and the triple-trace operators are a linear sum of the double-trace operators:
\begin{equation}\label{eq: su3tracetoy}
\begin{split}
&\Tr U^i \Tr U^j \Tr U^k=-\frac{2}{9}\Tr U^{i+j+k} +\frac{1}{3}\Tr U^i \Tr U^{j+k} +\frac{1}{3}\Tr U^{i+j} \Tr U^k +\frac{1}{3}\Tr ^{i+k} \Tr U^j \\
&+\frac{1}{3}\Tr U^{j-i} \Tr U^{k-i} -\frac{1}{9}\Tr U^{j+k-2i},\qquad\qquad \forall U \in SU(3)
\end{split}
\end{equation}

Here, the derivation of the double-trace loop equation stems from the condition that the shift in the probability measure must vanish ($n>0$):
\begin{equation}\label{eq: su3shift}
    \int  \mathcal{D} U\, \delta_\epsilon\left(  U^n_{ab} \Tr U^{m} \e^{\frac{N}{\lambda}( \tr U^\dagger +\tr U)}\right)=0
\end{equation}
with the definition of the variation:
\begin{equation}
    \delta_\epsilon (U_{ab})=U_{ab}+i\epsilon_{ac}U_{cb},\qquad\delta_\epsilon (U^\dagger_{ab})=U^\dagger_{ab}-iU^\dagger_{ac}\epsilon_{cb}
\end{equation}
Here $\epsilon$ is an arbitrary traceless Hermitian $3\times3$ matrix.  Expanding \eqref{eq: su3shift}, we get:
\begin{equation}
    \epsilon_{cd} \langle \frac{m}{N} U^n_{ab} U^{m}_{dc} +\sum_{i=0}^{n-1} U^i_{ac} U^{n-i}_{db}\Tr U^{m}+\frac{N}{\lambda} U^n_{ab} \Tr U^{m} (U_{dc}-U^\dagger_{dc})\rangle=0
\end{equation}
We observe that if $\epsilon_{cd} X_{dc} = 0$ for any traceless $\epsilon$, then it follows that $X_{dc} = \frac{1}{N} \delta_{dc} X_{ff}$, from which we can derive:
\eqsplit[]{
    &\langle \frac{m}{N} U^n_{ab} U^{m}_{dc} +\sum_{i=0}^{n-1} U^i_{ac} U^{n-i}_{db}\Tr U^{m}+\frac{N}{\lambda} U^n_{ab} \Tr U^{m} (U_{dc}-U^\dagger_{dc})\rangle\\
    &=\delta_{dc}\langle \frac{m}{N} U^n_{ab} \Tr U^{m} +\frac{n}{N} U^n_{ab}\Tr U^{m}+\frac{N}{\lambda} U^n_{ab} \Tr U^{m} (\Tr U-\Tr U^\dagger)\rangle
}
Upon contracting both sides with $\frac{1}{N^2} \delta_{ac}\delta_{bd}$, we arrive at the final form of the loop equation:
\eqsplit[eq: su3toyloopeq]{
&\langle \frac{m}{N^2} \Tr U^{n+m} +\sum_{i=0}^{n-1} \Tr U^i \Tr U^{n-i} \Tr U^{m}+\frac{1}{\lambda} \Tr U^{m} (\Tr U^{n+1}-\Tr U^{n-1})\rangle\\
&=\langle \frac{m+n}{N^2} \Tr U^n \Tr U^{m} +\frac{1}{\lambda} \Tr U^n \Tr U^{m} (\Tr U-\Tr U^\dagger)\rangle
}
The current double-trace loop equation is generally applicable to the SU(N) matrix model. For the U(N) matrix model, however, the left-hand side of the equation vanishes. By incorporating the trace identity \eqref{eq: su3tracetoy}, which is specific to SU(3), we get the loop equation that truncates at the double-trace variable level.

\subsection{Multi-trace Makeenko-Migdal loop equations}

In this section, we derive the general multi-trace Makeenko-Migdal loop equations for SU(N) and U(N) lattice gauge theories employing the Wilson action. The outcomes of this analysis, when integrated with the trace identities previously discussed, demonstrate that the loop equations for SU(N) lattice gauge theory close among $(N-1)$-trace variables. Conversely, for U(N) lattice gauge theory, closure occurs among $N$-trace variables.

The partition function for SU(N) and U(N) lattice Yang-Mills theories across different spacetime dimensions is given by:
\begin{equation}\label{eq: actionYM}
    Z = \int \mathcal{D} U \exp \left( -S \right), \quad \text{where} \,\, S = -\frac{N}{\lambda} \sum_p \tr U_p
\end{equation}
Here, $U_{p}$ represents the product of four group variables along four links that form a plaquette, with the summation extending over all orientations of all plaquettes. In two spacetime dimensions, our normalization ensures that the dynamics of the single plaquette Wilson loop, as specified by \eqref{eq: actionYM}, are equivalent to those in the one-matrix model \eqref{eq: onematrixsun}\footnote{Which does not apply of course to more complex loops in 2d, see~\cite{Kazakov:1980zj, Kazakov:1980zi, Kazakov:1981sb}.}. Furthermore, our normalization relates to $\beta$, as defined in \cite{Athenodorou:2021qvs}, by:
\begin{equation}
    \beta = \frac{2N^2}{\lambda}
\end{equation}

The multi-trace loop equation corresponds to the vanishing of the following variation:
\begin{equation}\label{eq: loopvar}
    0=\int \mathcal{D} U \delta_\epsilon\left(\mathcal{W}\left[\mathcal{C} (x, \mu, \nu)\right]_{ab}\prod_{i=1}^m \left(\Tr\mathcal{W}\left[\mathcal{C}_{i}\right]\right)  \exp \left(-S\right)\right).
\end{equation}
Here for the goal of deriving the loop equation, we are considering the following Wilson path:
\begin{equation}
    \mathcal{W}\left[\mathcal{C} (x, \mu, \nu)\right]_{ab}=\left( U_{x,\mu}...U_{y_-,\eta_-}U_{y,\eta}U_{y_+,\eta_+}...U_{x,\nu}^\dagger\right)_{ab}
\end{equation}
That is a specific Wilson path with open indices, starting at $x$ with $\mu$ direction and ending with $-\nu$ direction, at the same point $x$. Here $\delta_\epsilon$ is defined by taking the following variation on $U_{x,\mu}$\footnote{We notice that in this notation, $U_{x,\mu}^\dagger =U_{x+\mu,-\mu}$.}, whereas all the other link variables are unchanged:
\begin{equation}
    U_{x,\mu}\rightarrow (\mathds{I}+ i \epsilon) U_{x,\mu},\quad U_{x,\mu}^\dagger\rightarrow  U_{x,\mu}^\dagger (\mathds{I}- i \epsilon),\quad \epsilon^\dagger=\epsilon
\end{equation}
If we are talking about the SU(N) group instead of U(N), we need also to impose the tracelessness condition $\tr \epsilon=0$, which guarantees that the group element remains unimodular. 
Under this variation, we notice  that the action is changed by\footnote{To help set up our notation, we further explain that here:
\eqsplit[]{
&U_{p}(x,\mu,\nu)_{dc}=U_{x,\mu} U_{x+\mu,\nu}U_{x+\mu+\nu,-\mu}U_{x+\nu,-\nu}\\
&U_{p}(x,\mu,\nu)^\dagger_{dc}=U_{x,\nu} U_{x+\nu,\mu}U_{x+\mu,-\nu}^\dagger U_{x,\mu}^\dagger=U_{p}(x,\nu,\mu)_{dc}.
}}:
\begin{equation}
    \delta_\epsilon(S)=-i\frac{N}{\lambda}\epsilon_{cd}\sum_{\nu\neq \pm \mu}\left(U_{p}(x,\mu,\nu)-U_{p}(x,\mu,\nu)^\dagger\right)_{dc}
\end{equation}

Applying these variations to \eqref{eq: loopvar} we can get the loop equation in the following form:
\begin{equation}
    \epsilon_{cd}\langle A_{dc, ab} \rangle=0
\end{equation}
where $A_{dc, ab}$ reads:
\begin{small}
\begin{equation}
\begin{split}
&\prod_{i=1}^m \left(\Tr\mathcal{W}\left[\mathcal{C}_{i}\right]\right)\Bigg(\delta_{ac}\mathcal{W}\left[\mathcal{C} (x, \mu, \nu)\right]_{db}+\frac{N}{\lambda}\sum_{\eta\neq \pm \mu}\left(U_{p}(x,\mu,\eta)-U_{p}(x,\mu,\eta)^\dagger\right)_{dc}\mathcal{W}\left[\mathcal{C} (x, \mu, \nu)\right]_{ab}+\\
&\sum_{U_{y,\eta}=U_{x,\mu} }\mathcal{W}\left[\mathcal{C}_{y1} (x, \mu,-\mu_-)\right]_{ac}\mathcal{W}\left[\mathcal{C}_{y2} (x, \mu,\nu)\right]_{db}-\sum_{U_{y,\eta}=U_{x,\mu}^\dagger }\mathcal{W}\left[\mathcal{C}_{y1} (x, \mu,\mu)\right]_{ac}\mathcal{W}\left[\mathcal{C}_{y2} (x,\mu_+,\nu)\right]_{db}\Bigg)\\
&+\frac{1}{N}\left(\sum_{U_{y,\eta}=U_{x,\mu} }\mathcal{W}\left[\mathcal{C}_{n}(x,\mu, -\eta_-)\right]_{dc}-\sum_{U_{y,\eta}=U^\dagger_{x,\mu} }\mathcal{W}\left[\mathcal{C}_{n}(x,\eta_+, \mu)\right]_{dc}\right)\mathcal{W}\left[\mathcal{C} (x, \mu, \nu)\right]_{ab}\prod_{i=1,i\neq n}^m \left(\Tr\mathcal{W}\left[\mathcal{C}_{i}\right]\right)
\end{split}
\end{equation}
\end{small}

We further notice the fact that due to the arbitrariness of $\epsilon$, we have:
\begin{equation}\label{eq: tracetensor}
     \langle A_{dc, ab}\rangle=s_G \frac{\delta_{dc}}{N}\langle A_{ee, ab}\rangle
\end{equation}
Here $s_G$ is given by 
\begin{equation}\label{eq: sg}
        s_G= 
\begin{cases}
    1,& G=\text{SU(N)}\\
    0,              & G=\text{U(N)}
\end{cases}
\end{equation}
 which takes into account that  $\epsilon$ must be traceless for SU(N) gauge group.

 We further contract both sides of \eqref{eq: tracetensor} by $\frac{1}{N^2}\delta_{ac}\delta_{bd}$, and finally get the loop equation in the form:
\begin{equation}\label{eq: loop}
A_{\text{id}}+A_{\text{var}}+A_{\text{split}}+A_{\text{join}}=0
\end{equation}

\begin{figure}[t]
\begin{center}
\includegraphics[scale=0.4]{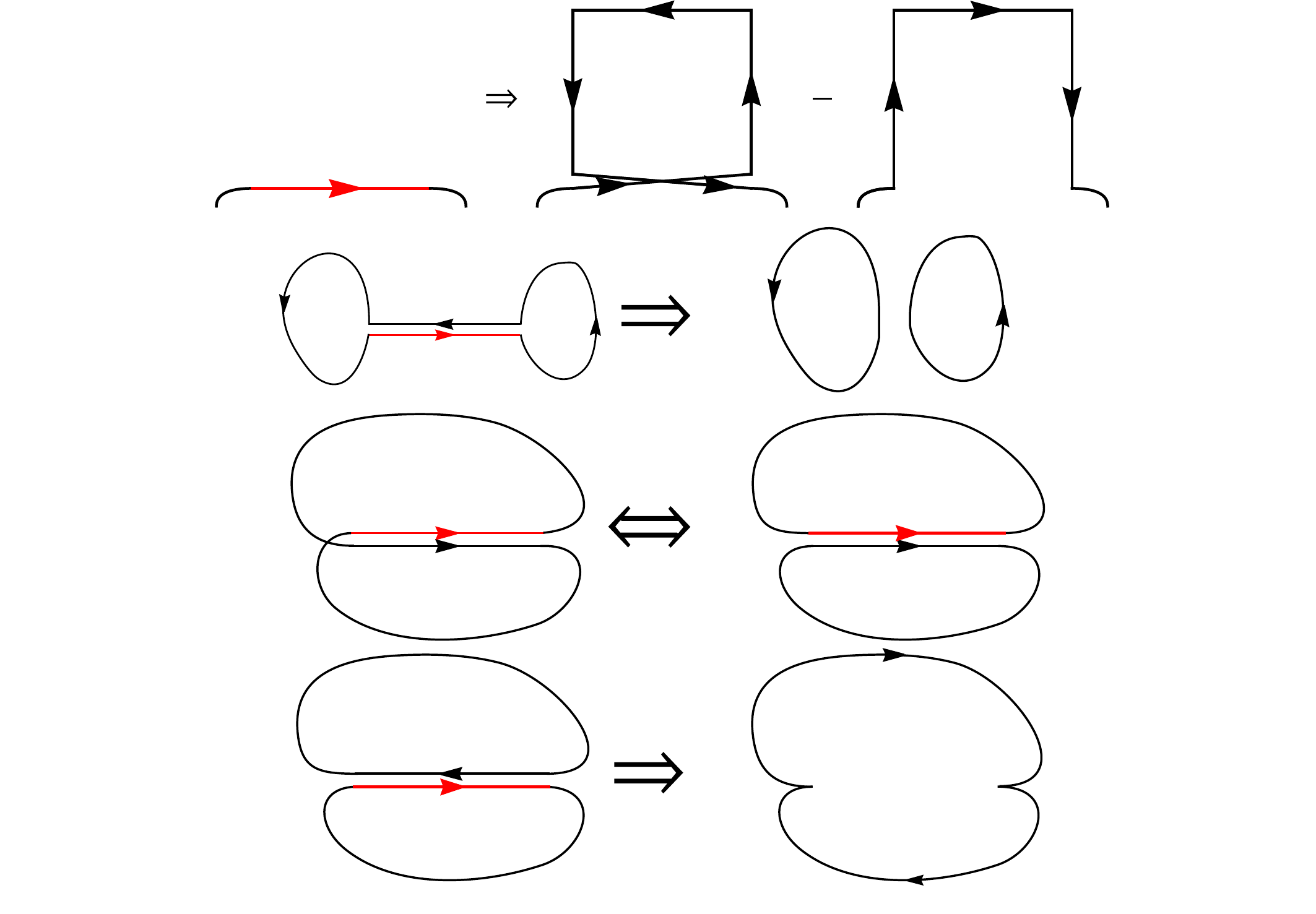}
\end{center}
\caption{Illustration of the different terms in the loop equation \eqref{eq: loop}.}  
\label{fig: loopvar}
\end{figure}
where we denoted:
\begin{equation}
    A_{\text{id}}=\left(1-s_G\frac{1+n_+-n_-+\sum_i^m(n_{i+}-n_{i-})}{N^2}\right)\langle
\Tr\mathcal{W}\left[\mathcal{C} (x, \mu, \nu)\right]\prod_{i=1}^m \Tr\mathcal{W}\left[\mathcal{C}_{i}\right]\rangle,
\end{equation}
\eqsplit[eq: var]{
A_{\text{var}}&=\frac{1}{\lambda}\langle\prod_{i=1}^m \Tr\mathcal{W}\left[\mathcal{C}_{i}\right]\sum_{\eta\neq \pm \mu}\left(\Tr\mathcal{W}\left[U_{p}(x,\mu,\eta)\mathcal{ C} \right]-\Tr\mathcal{W}\left[U_{p}(x,\mu,\eta)^\dagger\mathcal{C} \right]\right)\rangle\\
&-\frac{s_G}{\lambda}\langle\prod_{i=1}^m \Tr\mathcal{W}\left[\mathcal{C}_{i}\right]\sum_{\eta\neq \pm \mu}\left(\Tr\mathcal{W}\left[U_{p}(x,\mu,\eta)\right]\Tr\mathcal{W}\left[\mathcal{ C} \right]-\Tr\mathcal{W}\left[U_{p}(x,\mu,\eta)^\dagger\right]\Tr\mathcal{W}\left[\mathcal{C} \right]\right)\rangle
}
\begin{equation}
    A_{\text{join}}=\frac{1}{N^2}\langle\left(\sum_{U_{y,\eta}=U_{x,\mu} }\Tr\mathcal{W}\left[\mathcal{C}_{n}(x,\mu, -\eta_-)\circ \mathcal{C}\right]-\sum_{U_{y,\eta}=U^\dagger_{x,\mu} }\Tr\mathcal{W}\left[\mathcal{C}_{n}(x,\eta_+, \mu)\circ \mathcal{C} \right]\right)\prod_{i=1,i\neq n}^m \Tr\mathcal{W}\left[\mathcal{C}_{i}\right]\rangle,
\end{equation}
\begin{equation}
    A_{\text{split}}=\langle\prod_{i=1}^m \Tr\mathcal{W}\left[\mathcal{C}_{i}\right]\left(\sum_{U_{y,\eta}=U_{x,\mu} }\Tr\mathcal{W}\left[\mathcal{C}_{y1}\right]\Tr\mathcal{W}\left[\mathcal{C}_{y2}\right]-\sum_{U_{y,\eta}=U_{x,\mu}^\dagger }\Tr\mathcal{W}\left[\mathcal{C}_{y1} \right]\Tr\mathcal{W}\left[\mathcal{C}_{y2} \right]\right)\rangle.
\end{equation}
The term $A_{\text{id}}$ is proportional to the multi-trace Wilson loop under consideration. Here the terms \(n_+\) and \(n_-\) denote the number of overlaps with the link \(U_{x,\mu}\) in the contour \(\mathcal{C}\), depending on whether their direction aligns with \(\mu\). Similarly, \(n_{i,+}\) and \(n_{i,-}\) represent the number of overlaps with the link \(U_{x,\mu}\) in the contour \(\mathcal{C}_i\).  $A_{\text{var}}$ is the only part that is dynamical, in the sense that it depends on the content of the Wilson action. We notice that in the second line of \eqref{eq: var} vanishes for SU(2) due to the trace identities \eqref{eq: SU(2)trace}. $A_{\text{split}}$ and $A_{\text{join}}$ describe the splitting and joining of the Wilson loops. These terms are illustrated in the Figure~\ref{fig: loopvar}. We stress that the loop equation we derived so far is independent of the specific value of N. Practically we will set $m=N-2$ for SU(N) group and $m=N-1$ for U(N), substituting the finite N trace identities we discussed in Section~\ref{sec: trace} to the term $A_{\text{split}}$ and $A_{\text{var}}$, finally getting the loop equation that is closed on the $(m+1)$-trace Wilson loops.

For the specific case of SU(2) lattice gauge theory, which is extensively discussed in this article, we observe that the loop equation derived above can be further simplified. This simplification is achieved by substituting the trace identity for SU(2), as shown in \eqref{eq: SU(2)trace}:
\begin{align}\label{latticeSU(2)MME}
\frac{1}{\lambda}\sum_{\mu\neq \pm\nu} \Big[ W(\sigma^{l}_{\nu\mu}\circ C_{l_{\nu}})-W(\sigma^{l}_{\mu\nu}\circ C_{_{l_{\nu}}})\Big]+\frac{1}{2}\sum_{l'_{\nu}\sim l_{\nu}\in C} \,(-1)^{\frac{1+\nu\cdot\nu'}{2}}\left(W(C^{(1)}_{x,\nu}\circ \overline{C^{(2)}_{y,\nu'}})+\frac{1}{2}W(C)\right)=0\,   
\end{align}
Here, we specifically emphasize that the sum over \( l'_{\nu} \sim l_{\nu} \in C \) includes \( l_{\nu} \) itself.

\begin{figure}[t]
\begin{center}
\includegraphics[scale=0.4]{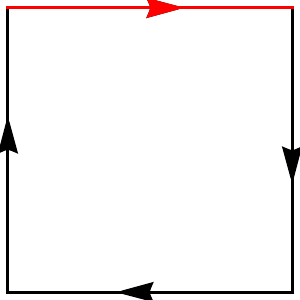}
\end{center}
\caption{The lowest order single-trace  loop equations.}  
\label{fig: loopeq}
\end{figure}

For illustration purposes, we present here the lowest-order single-trace loop equations in a two-dimensional lattice, which correspond to the variation of the Figure~\ref{fig: loopeq}:
\eqsplit[eq: 1eq]{
&\scalebox{1.6}{$0= \wthree-1+\wfour-\wtwo+(1-\frac{s_G}{N^2})\lambda\wone$}\\
&\quad\scalebox{1.6}{$-s_G$}\left(\doubletracetwo-\doubletraceone+\doubletracethree-\doubletracefour\right)
}

\subsubsection{Backtracks and redundancies}

\begin{figure}[t]
\begin{center}
\includegraphics[scale=0.4]{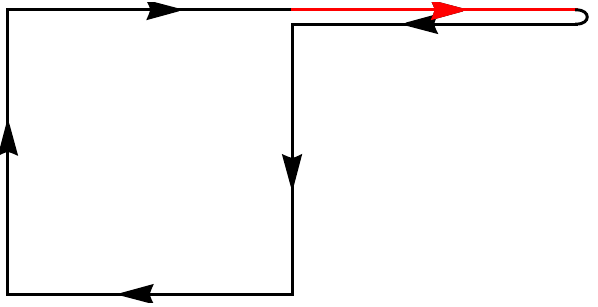}
\end{center}
\caption{Lowest order back-track loop equation.}  
\label{fig: backtrack}
\end{figure}
Backtracks play a rather subtle role in the formulation of the loop equations. They are defined by a string of Wilson lines (or even a backtrack ``tree" growing from a vertex of the contour)   that are equivalent to group identity due to unitarity. Backtrack loop equations originated from a variation on a ``cut" Wilson line, i.e. such that it has open indices inside it. Figure~\ref{fig: backtrack} shows the lowest order backtrack loop equation in a two-dimensional lattice:
\begin{equation}\label{eq: 2eq}
        \scalebox{1.6}{$\wtwo-\wfour-\wfive+\wsix=0$}
\end{equation}
More general situations contain a backtrack longer than the length one as shown in Figure~\ref{fig: backtrack}, the backtrack can also overlap with the non-backtrack part of the Wilson loop. In our study of SU(2) lattice Yang-Mills theory, approximately one-third to half of the loop equations are of the backtrack type. They play an essential role in constraining the space of (multi-trace) Wilson loops.

Another subtlety of the backtrack comes from the fact that it is equivalent to identity, so it can be inserted at any point of the Wilson loop. Combined with \eqref{eq: SU(2)trace}, we can expand the following double-trace Wilson loop to a combination of single-trace Wilson loops. Take as an example the double-trace Wilson loop corresponding to two loops:
\begin{equation}
    \doublezero
\end{equation}
There are many  different ways to insert here the backtracks and reduce them to a combination of single traces, e.g.:
\begin{equation}
    2\doubletwo= \redundantthree+\redundantfour
\end{equation}
\begin{equation}
    2\doubleone= \redundantone+\redundanttwo
\end{equation}
These insertions of the backtracks indicate that there are some kinematic redundancies in the space of Wilson loops. Similar kinematic redundancies also appear for double-trace Wilson loops for SU(3) lattice Yang-Mills theory. In addition to the redundancy arising from back-tracking, we observe inherent redundancies in the SU(3) trace identities \eqref{eq: SU(3)trace}. Given that the left-hand side of the equation remains invariant under permutations of $U$, $V$, and $W$, it is straightforward to apply permutations to the right-hand side. This process enables us to delineate the equations that identify redundancies among double-trace variables:
\begin{equation}
    \tr U^\dagger V \tr U^\dagger W -\tr U^\dagger V U^\dagger W=\tr V^\dagger U \tr V^\dagger W -\tr V^\dagger U V^\dagger W,\quad \forall U,V,W \in SU(3)
\end{equation}

\subsection{Makeenko-Migdal loop equations in the continuum}

In this section, we study the Makeenko-Migdal loop equations~\cite{Makeenko:1979pb} for the U(N) and SU(N) groups, within the context of Yang-Mills theory. The action for this theory is defined as:
\begin{equation}
   S=-\frac{1}{4g^2}\int d^{d}x\,\tr (F_{\mu\nu}F^{\mu\nu}) \text{, where } F_{\mu\nu}=\partial_{\nu}A_{\mu}-\partial_{\mu}A_{\nu}+[A_{\mu},A_{\nu}].
\end{equation}

Our focus will be on multi-trace Wilson loops, which are key observables in this theoretical framework:
\begin{align}\label{double-loop}
W(C_1,..., C_n)=\langle\Tr \mathcal{P}\exp \left(\oint_{C_1}  A_\mu dx^\mu\right)\,...\,\Tr \mathcal{P}\exp \left(\oint_{C_n}  A_\mu dx^\mu\right)\rangle\,.
\end{align}
While our primary emphasis will be on the single-trace and double-trace Wilson loops, the principles discussed can be extended to higher-trace loops.

We begin by examining the scenario for U(N) and subsequently explore the more intricate cases of SU(N), specifically SU(2) and SU(3). Our analysis will demonstrate that for these groups, the loop equations close respectively on single-trace and double-trace Wilson loops, illustrating the nuanced dynamics of these gauge theories.

From the fundamental principle that the functional integral of a full derivative vanishes\footnote{for the vanishing fields at space-time infinity as the boundary conditions}, we derive the \(U(N)\) Makeenko-Migdal Loop Equation, details of which are extensively discussed in the literature~\cite{Migdal:1983qrz,Gambini:1990dt, Makeenko:2023wvc}:
\begin{align}\label{MMLE}
\frac{\partial}{\partial x_\mu}\frac{\delta}{\delta\sigma_{\mu\nu}(x)}W(C_x)=\lambda \oint_{C}dy_{\nu}\,\delta^{(d)}(x-y)\,W(C^{(1)}_{xy}, C^{(2)}_{yx})
\end{align}
In this equation, \(\delta\sigma_{\mu\nu}\) represents the area derivative, a concept crucial to the formulation of loop equations in gauge theories. The operator \(\frac{\partial}{\partial x_\mu} \frac{\delta}{\delta\sigma_{\mu\nu}(x)}\) and further details on loop kinematics can be found in \cite{Migdal:1983qrz, Migdal:2022bka, Migdal:2023ppb}.

A simple generalization of  the single-loop MME allows to  derive the following U(N) loop equation for the double-loop Wilson average, in parallel to its lattice version \eqref{latticeSU(2)MME}:
\begin{align}\label{U(N)MMLEdouble}
&\frac{\partial}{\partial x_\mu}\frac{\delta}{\delta\sigma_{\mu\nu}(x)}W(C_x,C')=\notag\\
&=\lambda \oint_{C} dy_{\nu}\, \delta^{(d)}(x-y)\, W(C^{(1)}_{xy}, C^{(2)}_{yx},C') + \frac{\lambda}{N^2} \oint_{C} dy_{\nu}\, \delta^{(d)}(x-y)\, W(C_{xy} \circ C'_{yy})
\end{align}
In this equation, \(C_{xy} \circ C'_{yy}\) denotes the recomposition of two contours into a single loop at the point where they intersect, \(x=y\). This result illustrates that the U(N) loop equations naturally extend to encompass multi-trace configurations, effectively demonstrating that these equations close on N-trace Wilson loops.

Our exploration of the SU(N) loop equations begins with the observation that the U(N) Wilson average can be decomposed into two independent components when considering the decomposition \(U(N) \to SU(N) \times U(1)\):
\begin{align}\label{factorU(N)}
W(C) = \tilde W(C) \cdot w(C)    
\end{align}
Here, \(W(C)\) satisfies the Makeenko-Migdal loop equation as expressed in \eqref{MMLE}, and \(w\) adheres to the abelian loop equation:
\begin{align}\label{eq: MMabelian}
\frac{\partial}{\partial x_\mu}\frac{\delta}{\delta\sigma_{\mu\nu}(x)}w(C_x) = \frac{g^2}{N} \, w(C) \oint_{C} dy_{\nu} \, \delta^{(d)}(x-y)
\end{align}

Applying the area derivative, which acts as a linear operator, we leverage the geometric derivative \(\frac{\partial}{\partial x_\mu} \frac{\delta}{\delta\sigma_{\mu\nu}(x)}\) to both parts of the decomposed Wilson loop. This application yields the following combined equation for SU(N):
\begin{align}\label{SU(N)LE}
\frac{\partial}{\partial x_\mu}\frac{\delta}{\delta\sigma_{\mu\nu}(x)}W(C_x) = w(C) \frac{\partial}{\partial x_\mu}\frac{\delta}{\delta\sigma_{\mu\nu}(x)} \tilde W(C_x) + \frac{g^2}{N} \, \tilde W(C) w(C) \oint_{C} dy_{\nu} \, \delta^{(d)}(x-y)
\end{align}

Upon applying the U(N) loop equations \eqref{MMLE} and utilizing the factorization \eqref{factorU(N)}, we simplify the right-hand side (r.h.s.) and, by cancelling \(w(C)\) on both sides, derive the SU(N) loop equation for the single-loop Wilson average:

\begin{align}\label{SU(N)MMLE}
\frac{\partial}{\partial x_\mu}\frac{\delta}{\delta\sigma_{\mu\nu}(x)}\tilde W(C_x)=\lambda \oint_ {C}dy_{\nu}\,\delta^{(d)}(x-y)\,\,\left[\tilde W(C^{(1)}_{xy}, C^{(2)}_{yx})-\frac{1}{N^2}\tilde W(C)\right]
\end{align}

This equation indicates a refined understanding of the non-abelian contributions specific to SU(N) without the influence of the U(1) component.

Similarly, for the connected double-loop Wilson  average for U(N) gauge theory we also have the factorization:
\begin{align}\label{factorU(N)double}
W)_{U(N)}(C,C')=\tilde W_{SU(N)}(C,C')\cdot w_{U(1)}(C\circ C')  
\end{align}  
We apply the same procedure to the factorized double loop as we did for the single loop, leading to the following SU(N) double-loop loop equation:
\begin{align}\label{SU(N)MMLEdouble}
&\frac{\partial}{\partial x_\mu}\frac{\delta}{\delta\sigma_{\mu\nu}(x)}\tilde W(C_x,C')=\notag\\
&=\lambda\oint_{C}dy_{\nu}\,\delta^{(d)}(x-y)\,\,\left[\tilde W(C^{(1)}_{xy}, C^{(2)}_{yx},C')-\frac{1}{N^2}\tilde W(C,C')\right]+\notag\\ 
&+\lambda\oint_{C'}dy_{\nu}\,\delta^{(d)}(x-y)\Big[\,\,\tilde W(C_{xx}\circ C'_{yy})-\frac{1}{N^2}\tilde W(C,C')\Big]
\end{align}
Utilizing the trace identities elaborated in Section~\ref{sec: trace}, we arrive at the general conclusion that the SU(N) loop equations close on \((N-1)\)-trace variables. This principle significantly simplifies the mathematical framework for gauge theories by reducing the complexity of multi-loop interactions to more manageable forms.

Specifically for the \(SU(2)\) case, the application of the trace identity \eqref{eq: SU(2)trace}—expressed here as:
\begin{align}\label{su2Cid}
W\left(C_{1},C_{2}\right) = W\left(C_{1}\circ C_{2}\right) + W\left(C_{1}\circ \overline{C_{2}}\right),
\end{align}
transforms the loop equation \eqref{SU(N)MMLE} into a closed form within the single-loop space. The revised loop equation is:
\begin{align}\label{su2MMLE}
\frac{\partial}{\partial x_\mu}\frac{\delta}{\delta\sigma_{\mu\nu}(x)}W(C_x) = \frac{\lambda}{2} \oint_{C} dy_{\nu} \, \delta^{(d)}(x-y) \left[\,\,\frac{1}{2} W(C_x) + W\left(C^{(1)}_{xy} \circ \overline{C^{(2)}_{yx}}\right)\right]
\end{align}
In this equation, the notation \(\circ\) denotes the composition of two loops at open indices into a single-trace loop, and the overbar (\(\overline{\cdot}\)) indicates the inversion of the ordered product, akin to the Hermitian conjugation of the Wilson line.

For SU(3), the loop equations can similarly be closed on the double-loop space, thanks to the identity \eqref{eq: SU(3)trace}, which for this scenario is expressed as follows:
\eqsplit[]{
&W(C_{1},C_{2},C_{3})=\frac{1}{3}W(C_{1},C_{2}\circ C_{3})+\frac{1}{3}W(C_{2},C_{1}\circ C_{3})+\frac{1}{3}W(C_{3},C_{1}\circ C_{2})-\notag\\ 
&-\frac{1}{9}W(C_{1}\circ C_{2}\circ C_{3})-\frac{1}{9}W(C_{1}\circ C_{3}\circ C_{2})+\frac{1}{3}W(C_{1}\circ \overline{C_{3}},C_{2}\circ \overline{C_{3}})\\
&-\frac{1}{9}W(C_{1}\circ \overline{C_{3}}\circ C_{2}\circ \overline{C_{3}})
}

Applying this identity to the first term in the second line of \eqref{SU(N)MMLEdouble}, we derive a refined expression for the double-loop interaction:
\eqsplit[]{
&\tilde W(C^{(1)}_{xy}, C^{(2)}_{yx},C')=\frac{1}{3}\tilde W(C^{(1)}_{xy}, C^{(2)}_{yx}\circ C')+\frac{1}{3}\tilde W(C^{(2)}_{xy}, C^{(1)}_{yx}\circ C')+\frac{1}{3}\tilde W(C^{(1)}_{xy}\circ C^{(2)}_{yx},C')-\notag\\ 
&-\frac{1}{9}\tilde W(C^{(1)}_{xy}\circ C^{(2)}_{yx}\circ C')-\frac{1}{9}\tilde W(C^{(2)}_{xy}\circ C^{(1)}_{yx}\circ C')+\frac{1}{3}\tilde W(C^{(1)}_{xy}\circ\overline{ C'}, C^{(2)}_{yx}\circ\overline{ C'})\\
&-\frac{1}{9}\tilde W(C^{(1)}_{xy}\circ\overline{ C'}\circ C^{(2)}_{yx}\circ\overline{ C'})
}

Although contour \(C'\) does not necessarily intersect the point \(x=y\), we can conceptually connect it by introducing a backtrack path \(\Gamma_{xz} \otimes \Gamma_{zx}\), leading to the redefined interaction:
\begin{align}
  \tilde W(C^{(1)}_{xy}, C^{(2)}_{yx},C')\equiv  \tilde W(C^{(1)}_{xy}, C^{(2)}_{yx},\Gamma_{xz}\circ C_{zz}'\circ\Gamma_{zx})
\end{align}
Then, applying the identities outlined above, we can effectively recast the three-loop average into a combination of one- and two-loop terms, as shown on the right-hand side of the last identity. This substitution leads to the SU(3) loop equation for the Wilson averages, effectively closing the loop equations on the 1- and 2-loop space. Here, we will not make this substitution for the sake of brevity.
\newpage
\section{Positivity}\label{sec: positivity}

In the bootstrap method under consideration, the assertion of positivity is contingent upon the specific definition of the inner product within the operator space or subspace. More precisely, consider an expansion of operators in a given basis:
\begin{equation}
    \mathcal{O}=\sum_i \alpha_i \mathcal{O}_i
\end{equation}
Assuming a well-defined inner product in the operator space, the resulting matrix of the quadratic form of the scalar product is guaranteed to be positive semidefinite:
\begin{equation}\label{eq: inntosdp}
    \langle  \mathcal{O} | \mathcal{O}\rangle= \langle  \mathcal{O}^\dagger \mathcal{O}\rangle=  \alpha^{*\mathrm{T}} \mathcal{M}  \alpha\geq 0,\, \forall \alpha \Leftrightarrow \mathcal{M}\succeq 0.
\end{equation}
Here, the matrix \(\mathcal{M}\) is defined by the entries:
\begin{equation}
    \mathcal{M}_{ij}=\langle\mathcal{O}_i^\dagger \mathcal{O}_j\rangle
\end{equation}
Note that in this context, \(\dagger\) denotes the adjoint operators under the specified inner product.

\subsection{Hermitian conjugation}
This positivity condition is particularly evident when the measure is positive (i.e., devoid of a sign-problem). It is also the most well-understood category of positivity utilized in the bootstrap approach. In addition to the inherent positivity of the measure itself, this category encapsulates crucial information about the measure space~\cite{Kazakov:2021lel, Cho:2023ulr}. The resolution of the Hamburger moment problem exemplifies this principle.

Concretely, the positivity condition for this category is expressed as follows:
\begin{equation}
    \langle  \mathcal{O} | \mathcal{O}\rangle = \int d\mu \mathcal{O}^* \mathcal{O} \geq 0
\end{equation}
where $*$ denotes complex conjugation. Despite its seemingly straightforward appearance—as a result of integrating a positive function over a positive measure—this positivity condition can be remarkably potent, given its validity for any $\mathcal{O}$. The convergence of the bootstrap approach, predicated solely on the positivity of Hermitian conjugation, has been demonstrated in several model examples~\cite{Kazakov:2021lel, Cho:2023ulr}.
 
In the context of the matrix model, we articulate the following positivity condition:
\begin{equation}\label{eq: HerPos}
    \langle\frac{1}{N^m}(\sum_{\{O_i\}}\alpha_{\{O_i\}} O_{1, a_1b_1} \dots O_{m, a_mb_m})^\dagger(\sum_{\{O_i\}}\alpha_{\{O_i\}} O_{1, a_1b_1} \dots O_{m, a_mb_m})\rangle \geq 0
\end{equation}
where $O_i$ represents specific matrix operators and $a_i, b_i$ are their respective matrix indices, which are summed over. The coefficients list $\alpha_{\{O_i\}}$ is arbitrary here, as in \eqref{eq: inntosdp}. The condition \eqref{eq: HerPos} encapsulates $m$-traces operators. Intuitively, it appears sufficient to consider $m=N-1$ for the SU(N) matrix model and $m=N$ for both the U(N) and more generally the GL($N$) matrix models, as these configurations correspond to the truncation points of the loop equations. Further numerical analysis of the SU($N$) toy model \eqref{eq: onematrixsun} indicates that the imposition of $N$-traces positivity conditions does not introduce additional constraints beyond those imposed by $(N-1)$-traces positivity conditions. A more systematic exploration of this truncation phenomenon remains an open question.

As an example, for the SU(3) version of the toy model \eqref{eq: onematrixsun}, with a truncation $\Tr U^a \Tr U^b,\, |a|+|b|\leq 2$, we have the following positivity condition:
\begin{equation}\label{eq: toypossu3}
\begin{pNiceMatrix}[first-row,first-col]
      & U^\dagger_{a_1b_1} \mathbb{1}_{a_2b_2} &\mathbb{1}_{a_1b_1} U^\dagger_{a_2b_2} &\mathbb{1}_{a_1b_1} \mathbb{1}_{a_2b_2}&\mathbb{1}_{a_1b_1} U_{a_2b_2}  & U_{a_1b_1} \mathbb{1}_{a_2b_2} \\
U_{b_1a_1} \mathbb{1}_{b_2a_2}   & 1 & \left\langle \text{Tr} U^\dagger  \text{Tr} U \right\rangle  & \langle    \text{Tr} U \rangle  & \left\langle \text{Tr} U ^2\right\rangle  & \left\langle    \text{Tr} U^2 \right\rangle  \\
\mathbb{1}_{b_1a_1} U_{b_2a_2} & \left\langle \text{Tr} U^\dagger  \text{Tr} U \right\rangle  & 1 & \langle    \text{Tr} U \rangle  & \left\langle    \text{Tr} U^2 \right\rangle  & \left\langle \text{Tr} U ^2\right\rangle  \\
\mathbb{1}_{b_1a_1} \mathbb{1}_{b_2a_2}   & \langle    \text{Tr} U \rangle  & \langle    \text{Tr} U \rangle  & 1 & \langle    \text{Tr} U \rangle  & \langle    \text{Tr} U \rangle  \\
\mathbb{1}_{b_1a_1} U^\dagger_{b_2a_2}   & \left\langle \text{Tr} U ^2\right\rangle  & \left\langle    \text{Tr} U^2 \right\rangle  & \langle    \text{Tr} U \rangle  & 1 & \left\langle \text{Tr} U^\dagger  \text{Tr} U \right\rangle  \\
U^\dagger_{b_1a_1} \mathbb{1}_{b_2a_2}   & \left\langle    \text{Tr} U^2 \right\rangle  & \left\langle \text{Tr} U ^2\right\rangle  & \langle    \text{Tr} U \rangle  & \left\langle \text{Tr} U^\dagger  \text{Tr} U \right\rangle  & 1 \\
\end{pNiceMatrix}\succeq 0.
\end{equation}

A more intricate example can be found in the following scenario, which outlines the positivity condition for the 2-dimensional lattice Yang-Mills theory. Here, all open Wilson lines initiate and terminate at the origin, marked by a red dot, and are subject to a length cutoff~$=4$. This configuration corresponds to a $9 \times 9$ matrix that includes 6 independent Wilson loops:

\begin{equation}\label{eq: posym2d}
    \begin{pNiceMatrix}[first-row,first-col]
    & \vcenter{\hbox{$\mathbb{1}$}} & \lineone& \linetwo& \linethree& \linefour& \linefive& \linesix& \lineseven& \lineeight & \\
   \vcenter{\hbox{$\mathbb{1}\quad$}} & 1 & \wone & \wone & \wone & \wone & \wone & \wone & \wone & \wone \\
   \linefive & \wone & 1 & \wtwo & \wtwo & \wsix & \wthree & \wfour & \wfour & \wfive \\
   \lineseven& \wone & \wtwo & 1 & \wsix & \wtwo & \wfour & \wfive & \wthree & \wfour \\
   \linesix & \wone & \wtwo & \wsix & 1 & \wtwo & \wfour & \wthree & \wfive & \wfour \\
   \lineeight & \wone & \wsix & \wtwo & \wtwo & 1 & \wfive & \wfour & \wfour & \wthree \\
   \lineone &\wone & \wthree & \wfour & \wfour & \wfive & 1 & \wtwo & \wtwo & \wsix \\
   \linethree & \wone & \wfour & \wfive & \wthree & \wfour & \wtwo & 1 & \wsix & \wtwo \\
   \linetwo & \wone & \wfour & \wthree & \wfive & \wfour & \wtwo & \wsix & 1 & \wtwo \\
   \linefour & \wone & \wfive & \wfour & \wfour & \wthree & \wsix & \wtwo & \wtwo & 1 \\ 
 \end{pNiceMatrix}\succeq 0.
\end{equation}

\subsection{Reflection Positivity}

The study of reflection positivity was initiated in the context of constructive quantum field theory \cite{Osterwalder:1973dx, Osterwalder:1974tc}, where it comes as part of the condition to guarantee that a Euclidean field theory defines a consistent relativistic quantum field theory through Wick rotation. The conformal bootstrap\cite{Belavin:1984vu, Rattazzi:2008pe}\footnote{For recent developments in this field, see \cite{Poland:2018epd} and \cite{Rychkov:2023wsd}, which provide comprehensive updates and insights into current research trends.} and the infrared bound\cite{Frohlich:1976xk} are among the most fruitful applications of reflection positivity. 

\begin{figure}[t]
\begin{center}
\includegraphics[scale=0.6]{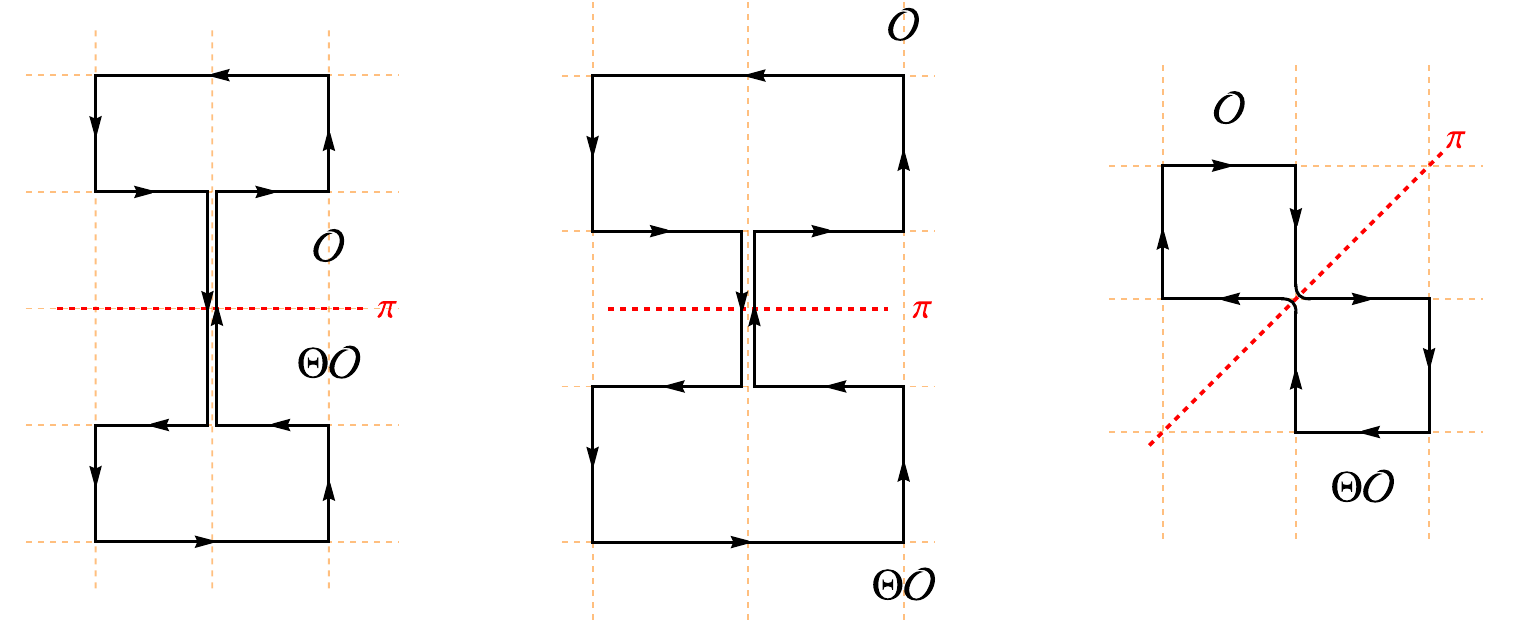}
\end{center}
\caption{Here are examples of site-reflection, link-reflection, and diagonal-reflection. It is important to note that the intersection of the Wilson path with the reflection plane $\pi$ does not necessarily occur at the same point, as illustrated in the figure.}  \label{fig: ref}
\end{figure}

As illustrated in Fig~\ref{fig: ref}, it is widely accepted that there are three known types of reflection positivity in lattice Yang-Mills theory on the square lattice: site-reflection, link-reflection\cite{Osterwalder:1977pc}, and diagonal-reflection\cite{Kazakov:2022xuh}. The inner product defined by the reflection is straightforward in the subspace of operators located on the positive side of the reflection plane:\footnote{The positive side of the reflection plane can be chosen arbitrarily. It is marked by ``+" superscript for the corresponding operators.}
\begin{equation}
    \langle\mathcal{O}^+|\mathcal{O}^+\rangle=\langle\mathcal{O}^+\Theta \mathcal{O}^+\rangle
\end{equation}
In the infrared (IR) regime of the theory, this same inner product undergoes a transformation via Wick rotation to become the inner product of the physical Hilbert space\cite{Osterwalder:1977pc}.

\begin{figure}[t]
\begin{center}
\includegraphics[scale=0.5]{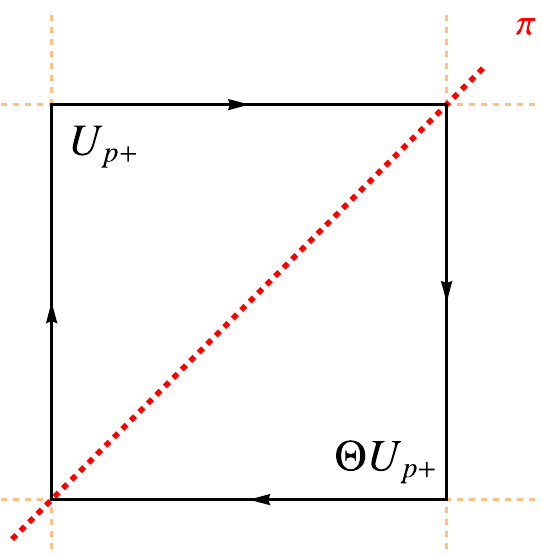}
\end{center}
\caption{ An example of plaquette who intersect with the reflection plane $\pi$.}  \label{fig: upref}
\end{figure}

As an example, below is a closer investigation of the diagonal reflection positivity. It treats the diagonal of the square lattice as the surface of quantization. The reflection operation defined on a lattice site with lattice coordinates \((x_1, x_2, \ldots, x_n)\) is formalized as follows:
\begin{equation}
    \Theta (x_1, x_2, \ldots, x_n) = (x_2, x_1, \ldots, x_n)
\end{equation}
This operation swaps the first two coordinates while leaving the remaining coordinates unchanged, reflecting the site across the line or plane that interchanges these dimensions.

For a unitary matrix associated with a link on the lattice, the reflection transformation is defined as follows:
\begin{equation}
    \Theta U_{x \rightarrow y} = U_{\Theta x \rightarrow \Theta y}^\dagger
\end{equation}

The proof of the reflection positivity is straightforward. We begin by noticing the following fact:
\begin{equation}
    \int \mathcal{D} U \mathcal{O}^+\Theta \mathcal{O}^+=\int \mathcal{D} U^+ \mathcal{O}^+\int \mathcal{D} U^\dagger \Theta \mathcal{O}^+=\parallel\int \mathcal{D} U^+ \mathcal{O}^+\parallel^2\geq 0
\end{equation}
Here $\mathcal{O}^+$ can be arbitrary operators located in the positive side of the reflection plane, or in our case as in Fig~\ref{fig: upref}: $x_2>x_1$.

To consider the expectation value with the Wilson action, we notice that the main obstacle is the terms $S_\Delta$ in the action that are represented by the plaquettes intersected (along their diagonals, as in Fig.~\ref{fig: upref}) by the reflection plane $\pi$:
\eqsplit[eq: refpos]{
&\int \mathcal{D} U \mathcal{O}^+\Theta \mathcal{O}^+ \exp(-S)=\int \mathcal{D} U \mathcal{O}^+\Theta \mathcal{O}^+ \exp(-S_+-\Theta S_- +S_\Delta)\\
&=\int \mathcal{D} U \mathcal{O}^+\exp(-S_+)\Theta(\mathcal{O}^+\exp(-S_+)) \exp(-S_\Delta)
}

But this is appears not a  problem once we  notice that:
\begin{equation}
    \exp(-S_\Delta)=\prod_{p\in\Delta} \exp(\frac{N}{\lambda} \Tr U_p)=\prod_{p\in\Delta} \exp(\frac{N}{\lambda} \Tr U_p^+\Theta U_p^+)
\end{equation}
Here the product of $p\in\Delta$ means the product of all the plaquettes intersecting the reflection plane $\pi$. Expanding the exponent and the infinite product, we see that this term is a positive sum of terms of the form $\mathcal{O}^+\Theta \mathcal{O}^+$, which means that the whole expression  \eqref{eq: refpos} is positive.

\subsection{Symmetry Reduction}

The primary advantage of the bootstrap approach is that it naturally takes into account the symmetries of the theory. This feature has been observed in such subjects as the conformal bootstrap and S-matrix bootstrap~\cite{Rattazzi:2010yc, He:2018uxa, Cordova:2018uop, Paulos:2018fym}. In the context of the bootstrap approach discussed in this project, the symmetries of the model under consideration are often  manifested in the inner products defined as follows:
\begin{equation}
    \langle (g \circ \mathcal{O}_1) | (g \circ \mathcal{O}_2) \rangle = \langle \mathcal{O}_1 | \mathcal{O}_2 \rangle, \forall g \in G
\end{equation}
Here, $g \circ \mathcal{O}$ denotes the transformation of the operator $\mathcal{O}$ under the symmetry group $G$, typically representing a reducible representation of this group.

Thanks to Schur's lemma, we know that due to group-theoretical reasons, the operators belonging to different irreducible representations of the symmetry group must be orthogonal. Consequently, the positive semidefinite matrix defined in \eqref{eq: inntosdp} becomes block diagonal, where each block corresponds to an irreducible representation. However, these blocks can be further decomposed using the methods of Invariant Semidefinite Programming\footnote{For example, the interested reader could refer to \cite{bachoc2012invariant}.}. More specifically, suppose the space of operators forms a reducible representation of the symmetry group $G$ with the following expansion into irreducible representations:
\begin{equation}
    V=\bigoplus_{k=1}^D \mathrm{Rep}_k^{\bigoplus m_k},
\end{equation}
where $m_k$ is the multiplicity of a given irreducible representation $\text{Rep}_k$. The block corresponding to this irreducible representation then has dimensions $m_k \times m_k$. It is noteworthy that the final dimension of the positivity matrix is $\sum_k m_k$, rather than $\text{dim}V = \sum_k d_k m_k$.\footnote{This can be understood from the classical example of a Hamiltonian with SU(2) symmetry, where states with the same $l$ but different $m$ possess identical energy eigenvalues. In such cases, the lesson is that positivity among the highest weight vectors already captures all necessary information regarding the positivity.}

Implementing this symmetry reduction is analogous to projecting the physical state with respect to spin and parity in conformal or $S$-matrix bootstrap. The systematic steps to obtain the projector are outlined as follows \cite{Kazakov:2021lel}:
\begin{enumerate}
    \item Identify a specific realization of every irreducible representation (irrep) of the invariant group using GAP software~\cite{GAP4}.
    \item Apply the algorithm proposed in \cite{xu2021computation} to find an equivalent real representation, if the irrep provided by GAP is complex.
    \item Decompose into such irreps using the projector to \(\mathrm{Rep}_k\) ~\cite{serre1977linear}:
\begin{equation}
    p_{\alpha\alpha, k} = \frac{\dim (\mathrm{Rep}_k)}{\dim G}\sum_{g \in G} r_{\alpha \alpha}(g^{-1}) g
\end{equation}
Here, \(r_{\alpha \beta}\) represents the matrix elements of a real representation identified at step 2, with \(\alpha,\, \beta = 1, 2, \dots, \dim(\mathrm{Rep}_k)\). By setting \(\alpha = 1\), \(P_k = p_{11, k}\) serves as a projector to \(\mathrm{Rep}_k\).
\end{enumerate}

Some remarks follow:
\begin{itemize}
    \item At the second step, the preference for a real representation stems from the advantage that real Semidefinite-Programming typically offers big advantages w.r.t. complex Semidefinite-Programming\footnote{We note that all multi-trace Wilson loops are real, a property stemming from the unimodularity of the corresponding Haar measure.}. Fortuitously, for the lattice Yang-Mills theory under investigation, all the groups possess real representations. A systematic method to determine whether a finite group has real irreducible representations is documented in \cite{serre1977linear}.
    \item At the third step, selecting \(\alpha=1\) for the final projector might appear arbitrary. However, the key insight is that choosing different \(\alpha\) values yields SDP blocks that are fully equivalent to those obtained with \(\alpha=1\).
\end{itemize}

\begin{table}
\begin{center}
    \begin{tabular}{ |c ||p{2cm}|p{2cm}|p{2cm}|  }
 \hline
 Dimension& Hermitian Conjugation &site\&link reflection &diagonal reflection\\
 \hline
 2& \(B_2\times \mathbb{Z}_2\) & \(\mathbb{Z}_2\times \mathbb{Z}_2\) & \(\mathbb{Z}_2\times \mathbb{Z}_2\)   \\
 3& \(B_3\times \mathbb{Z}_2\) & \(B_2\times \mathbb{Z}_2\) & \(\mathbb{Z}_2^3\)  \\
 4& \(B_4\times \mathbb{Z}_2\) & \(B_3\times \mathbb{Z}_2\) & \(B_2\times \mathbb{Z}_2^2\)\\
 \hline
\end{tabular}
\caption{Invariant groups of correlation and reflection matrices.}
\label{tab: latticesymmetry}
\end{center}
\end{table}

For the model under consideration, specifically SU(N) or U(N) lattice Yang-Mills theory, we observe the global symmetries listed in Table~\ref{tab: latticesymmetry}. The notation for the group $B_n$ is adopted from the nomenclature used on the Wikipedia page for the Hyperoctahedral Group (\href{https://en.wikipedia.org/wiki/Hyperoctahedral_group}{Hyperoctahedral Group}). In more standard terms, $B_2$ and $B_3$ correspond to the dihedral and octahedral groups, respectively:
\begin{equation}
    B_2 \simeq D_4, \quad B_3 \simeq O_h
\end{equation}

The $\mathbb{Z}_2$ in Table~\ref{tab: latticesymmetry} represents the charge conjugation $C$, which acts by inverting the direction of a Wilson loop. 

The discussion above parallels the considerations in the planar limit as presented by the author\cite{Kazakov:2022xuh}. However, in the case of multi-trace operators, an additional symmetry emerges—specifically: the permutation group among different traces. Consequently, there is an \(S_m\) group that complements the global symmetry group listed in Table~\ref{tab: latticesymmetry}.

To illustrate this principle, we conduct the symmetry reduction of \eqref{eq: toypossu3}. Besides the permutation group previously discussed, the corresponding model \eqref{eq: onematrixsun} also exhibits charge conjugation symmetry, where \(U \rightarrow U^\dagger\). Therefore, we can categorize the operators according to their representations under the \(\mathbb{Z}_2^2\) group:
\eqsplit[]{
(+,+):&\quad \mathbb{1},\quad U^\dagger_{a_1b_1} \mathbb{1}_{a_2b_2} +\mathbb{1}_{a_1b_1} U^\dagger_{a_2b_2} +\mathbb{1}_{a_1b_1} U_{a_2b_2}  + U_{a_1b_1} \mathbb{1}_{a_2b_2}\\
(-,+):&\quad \quad (U^\dagger_{a_1b_1} \mathbb{1}_{a_2b_2} +\mathbb{1}_{a_1b_1} U^\dagger_{a_2b_2}) -(\mathbb{1}_{a_1b_1} U_{a_2b_2}  + U_{a_1b_1} \mathbb{1}_{a_2b_2})\\
(+,-):&\quad \quad (U^\dagger_{a_1b_1} \mathbb{1}_{a_2b_2} -\mathbb{1}_{a_1b_1} U^\dagger_{a_2b_2}) +(\mathbb{1}_{a_1b_1} U_{a_2b_2}  - U_{a_1b_1} \mathbb{1}_{a_2b_2})\\
(-,-):&\quad \quad (U^\dagger_{a_1b_1} \mathbb{1}_{a_2b_2} -\mathbb{1}_{a_1b_1} U^\dagger_{a_2b_2}) -(\mathbb{1}_{a_1b_1} U_{a_2b_2}  - U_{a_1b_1} \mathbb{1}_{a_2b_2})
} 

Consequently, the positivity condition detailed in \eqref{eq: toypossu3} is further refined to the following specifications:
\eqsplit[]{
\begin{pmatrix}
1 & \langle\Tr U\rangle \\
\langle\Tr U\rangle & \frac{1}{4} \langle\Tr U\Tr U^\dagger\rangle+\frac{1}{4} \langle\Tr U\Tr U\rangle+\frac{1}{4} \langle\Tr U^2\rangle+\frac{1}{4} 
\end{pmatrix}& \succeq 0\,,\\
\frac{1}{4} \langle\Tr U\Tr U^\dagger\rangle-\frac{1}{4} \langle\Tr U\Tr U\rangle-\frac{1}{4} \langle\Tr U^2\rangle+\frac{1}{4} &\geq 0\\
-\frac{1}{4} \langle\Tr U\Tr U^\dagger\rangle-\frac{1}{4} \langle\Tr U\Tr U\rangle+\frac{1}{4} \langle\Tr U^2\rangle+\frac{1}{4}&\geq 0\\
-\frac{1}{4} \langle\Tr U\Tr U^\dagger\rangle+\frac{1}{4} \langle\Tr U\Tr U\rangle-\frac{1}{4} \langle\Tr U^2\rangle+\frac{1}{4}&\geq 0
}
The symmetry reduction of the matrix described in \eqref{eq: posym2d} has been thoroughly detailed in the supplementary material of \cite{Kazakov:2022xuh}. To avoid redundancy, we will not repeat the discussion here.

In concluding this discussion, it is important to note that while our analysis has primarily focused on finite groups, the results are readily generalizable to Lie groups, particularly compact Lie groups\cite{BFSSWIP}. This extension is supported by established correspondences between compact Lie groups and finite groups, suggesting that our findings may have broader applications across various mathematical structures in theoretical physics. This potential for generalization not only underscores the versatility of our theoretical framework but also highlights its relevance for ongoing research in the field of quantum mechanics and field theory, where Lie groups play a crucial role.
 
\newpage
\section{Bootstrap procedure}

Our previous discussion of the loop equations in Section~\ref{sec: loopeq} and the positivity conditions in Section~\ref{sec: positivity} facilitate the straightforward bounding of several observables within the corresponding theory. In the subsequent subsection, we will present some immediate results for the bootstrap. However, for the lattice Yang-Mills theory, formulating a large-scale bootstrap problem is not as straightforward as it might initially appear. The primary challenge lies in the selection of operators for our bootstrap analysis. In Section~\ref{sec: hierachy}, we propose a hierarchy that assigns each operator (open Wilson line) and loop variable (Wilson loop) a specific level, and truncates the infinite space of Wilson lines according to its level. At last, we present the results we have achieved so far, mostly concerned with the free energy per plaquette and string tension, which we compare with the existing Monte Carlo calculations, as well as strong and weak coupling expansions.

\subsection{Initiate the bootstrap}
\begin{figure}
\centering
\begin{subfigure}{.5\textwidth}
  \centering
  \includegraphics[width=\linewidth]{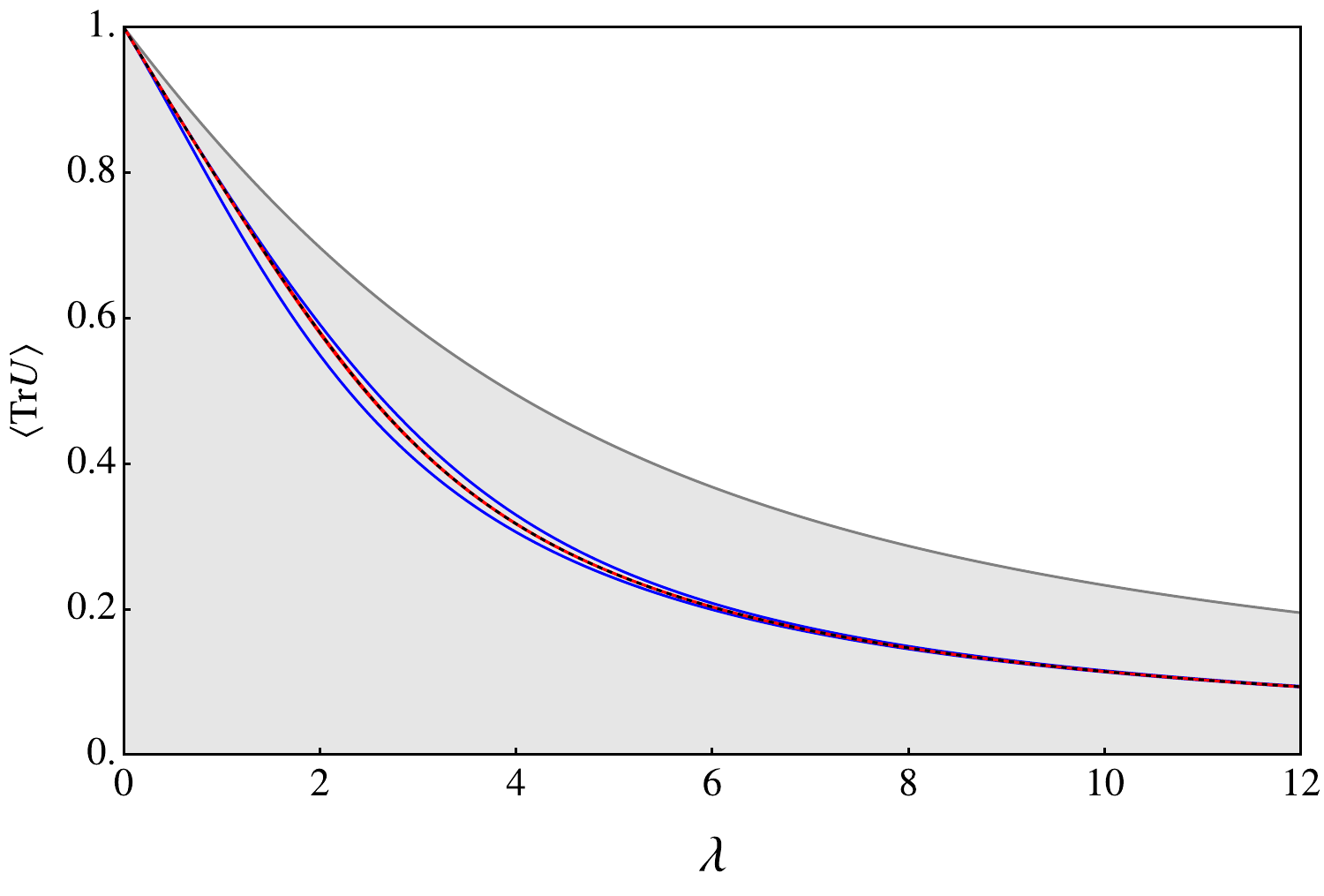}
\end{subfigure}%
\begin{subfigure}{.5\textwidth}
  \centering
  \includegraphics[width=\linewidth]{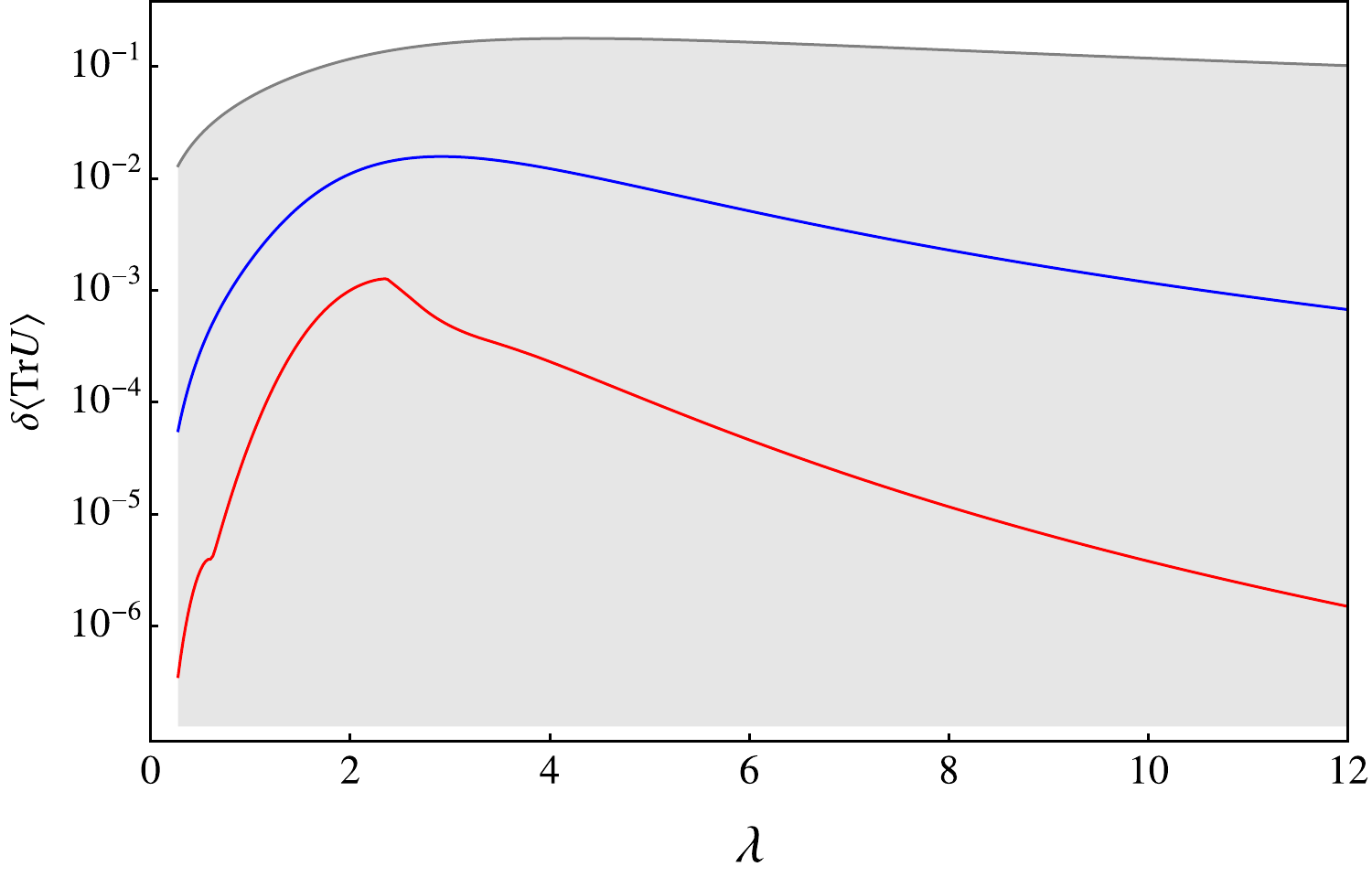}
\end{subfigure}
\caption{This figure presents the bounds for the SU(3) one-matrix model. On the left side of the figure, we depict both the upper and lower bounds for $\langle \Tr U \rangle$ at different values of $\Lambda$: 1 (gray), 2 (blue), and 3 (red). The black dashed line represents the exact value of the model. On the right side of the figure, we illustrate the difference between the upper bound and the exact value on a logarithmic scale, which demonstrates a clear exponential convergence.
}
\label{fig: toysu3}
\end{figure}

The bootstrap of the one-matrix model \eqref{eq: onematrixsun} is straightforwardly implemented. Here, we present the SU(3) case of equation \eqref{eq: onematrixsun} as an example\footnote{In Appendix~\ref{app: moreboot}, we also present the bootstrap results for SU(2), U(2), and U($\infty$).}. The fundamental components are already summarized in Section~\ref{sec: loopeq} and Section~\ref{sec: positivity}. The set of loop equations combines those from \eqref{eq: su3toyloopeq} with the trace identities from \eqref{eq: su3tracetoy}. To solve this system, the \textbf{Reduce} function in Mathematica is employed, revealing that all double-trace variables are linear functions of two variables: $\langle\Tr U\rangle$ and $\langle\Tr U\Tr U^\dagger\rangle$.

Concerning the positivity condition, as exemplified in \eqref{eq: toypossu3}, it arises from the inner product of double trace operators \(U^m_{a_1b_1} U^n_{a_2b_2}\). We truncate the space of operators such that \(|m| + |n| \leq \Lambda\)\footnote{The positivity condition \eqref{eq: toypossu3} corresponds to \(\Lambda=1\).}. Increasing \(\Lambda\) yields progressively tighter dynamical bounds on the observables.

The results of the bootstrap up to \(\Lambda=3\) are illustrated in Figure~\ref{fig: toysu3}. Both the upper and lower bounds are observed converging towards the exact value\footnote{The exact value for the current one-matrix model is documented in Appendix~\ref{app: exact}. For illustrative Mathematica code used to generate the plot in Figure~\ref{fig: toysu3} and for numerical examples referenced in Appendix~\ref{app: moreboot}, please visit the GitHub page \href{https://github.com/Canonical111/Unitary-Matrix-Model-Bootstrap-Example}{https://github.com/Canonical111/Unitary-Matrix-Model-Bootstrap-Example}.}. In the right panel of Figure~\ref{fig: toysu3}, the convergence of the upper bound to the exact value for \(\Lambda=1, 2, 3\) is depicted, demonstrating a clear exponential trend.

Before exploring the details of the hierarchy discussed in the next section, it is important to note that from the equations and positivity conditions available for the lattice Yang-Mills theory, we can already derive non-trivial upper bounds on the observables. Specifically, for SU(2) and general U(N) models, we integrate the loop equations \eqref{eq: 1eq} and \eqref{eq: 2eq} with the positivity condition \eqref{eq: posym2d}. This setup constitutes a bootstrap problem involving six variables, which encompass all variables below level 2 in the hierarchy discussed in the subsequent section. By minimizing the plaquette average, we obtain non-trivial bounds for SU(2) and uniform bounds for U(N) (including U($\infty$)). For instance, at \(\lambda = 2\), the bounds are:
\begin{equation}
    \langle\Tr U\rangle \leq \begin{cases} 
      0.693 & \text{for U(N)}\\
      0.75755 & \text{for SU(2)} 
   \end{cases}
\end{equation}

\subsection{Hierachy}\label{sec: hierachy}

Our framework for the bootstrap program applied to the finite \(N\) matrix models and lattice Yang-Mills theories can be summarized as follows:
\begin{equation}\label{eq: dual}
\begin{aligned}
\min/\max \quad & \wone,\\
\textrm{subject to} \quad & \text{MM loop equations},\\
& \text{HerM}^{\text{irrep}} \succeq 0,\\
& \text{RefM}^{\text{irrep}} \succeq 0, \times 3
\end{aligned}
\end{equation}
Here, the objective function can, in principle, minimize any convex function of the Wilson loops. We enforce the (multi-trace) Makeenko-Migdal loop equations and apply trace identities to eliminate higher multi-trace Wilson loops. Additionally, four types of positivity conditions are imposed, three of which are reflection positivities, each factorizable into different irreducible representations of the corresponding symmetry group. In the case of large \(N\), we also implemented in  \cite{Kazakov:2021lel, Kazakov:2022xuh}, in addition to these constraints,  a relaxation of the quadratic variables and ensure the positivity of the relaxation matrix \(\mathcal{R} \succeq 0\).

One critical feature remains to add: to effectively formulate the bootstrap problem \eqref{eq: dual}, we must implement a finite truncation of the Wilson loops. This truncation is essential for the efficiency of the bootstrap process. Below, we will illustrate this truncation method (hierarchy) using lattice Yang-Mills theory as an example. We believe that this principle can be universally applied to bootstrap methods within this category of problems.

Our practical selection of positivity generalizes \eqref{eq: posym2d} by focusing only on the positivity of the inner products of Wilson lines that initiate and terminate at the origin point. This selection is based on the fact that such a subset of Wilson lines forms a representation of the largest possible symmetry group. We then derive all necessary equations among the Wilson loops within the matrix for the inner products of these Wilson lines.

For the selection of Wilson lines as the basis of the inner product, we employ a hierarchical system, where each Wilson line starting and ending at zero is assigned an integer number representing its level. This hierarchy is defined recursively: we begin with the identity operator at level \(0\) and a single plaquette at level \(1\). Subsequent operators are derived from the Schwinger-Dyson variations, as illustrated in the first line of Figure~\ref{fig: loopvar}. For instance, in the context of SU(2) lattice gauge theory in two dimensions, the following defines a level 2 Wilson line:
\begin{equation}
    \levelone,\, \leveltwo,\, \levelthree,\, \levelfour,\, \levelfive
\end{equation}
The red dot in our diagrams represents the origin point, where the Wilson line has an open index. We also define the hierarchy of Wilson loop expectations, which is determined by the sum of the levels of the Wilson lines whose inner products generate these loops. It is important to note that if multiple levels could be defined recursively for a Wilson line/loop, the smallest one is selected as its level.

As an illustrative example, we list all level two Wilson loops for the two-dimensional SU(2) lattice gauge theory:
\begin{equation}
  \scalebox{1.8}{$\wtwo,\,\wthree,\,\wfour,\,\wfive,\,\wsix,\,$}
\end{equation}
These loops, along with the single plaquette Wilson loop, comprise all the variables that appear in the matrix \eqref{eq: posym2d}.

This hierarchically defined structure ensures that the variables in our positivity conditions are interrelated through the loop equations. When constructing the inner product matrix, if we consider Wilson lines up to level $\Lambda$, then the longest Wilson loop in our problem will be $8 \times \Lambda$. In the current project, we typically consider up to $\Lambda=3$, corresponding to a maximum loop length of 24.
\begin{figure}[t]
\centering
\begin{subfigure}{.5\textwidth}
  \centering
  \includegraphics[width=\linewidth]{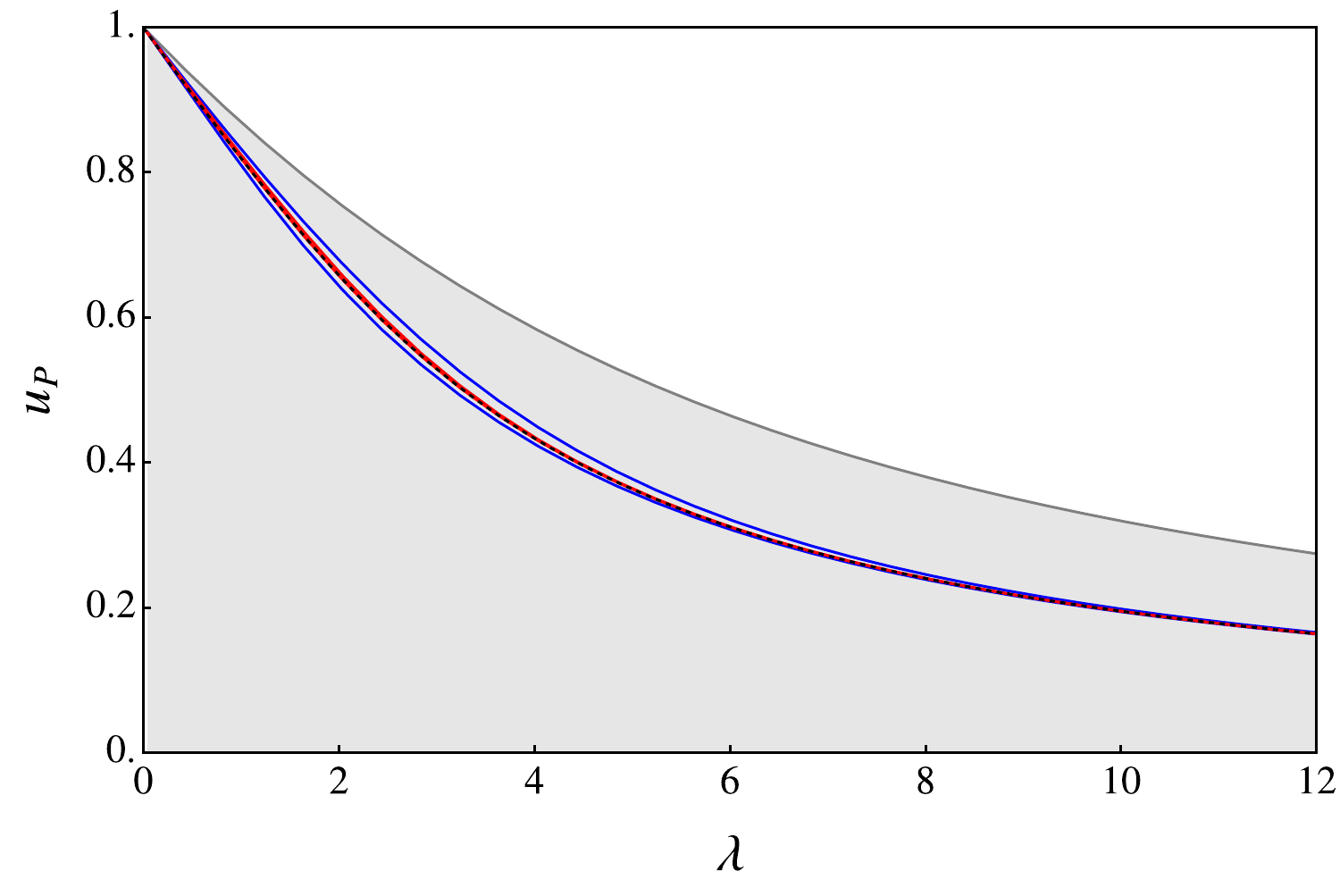}
\end{subfigure}%
\begin{subfigure}{.5\textwidth}
  \centering
  \includegraphics[width=\linewidth]{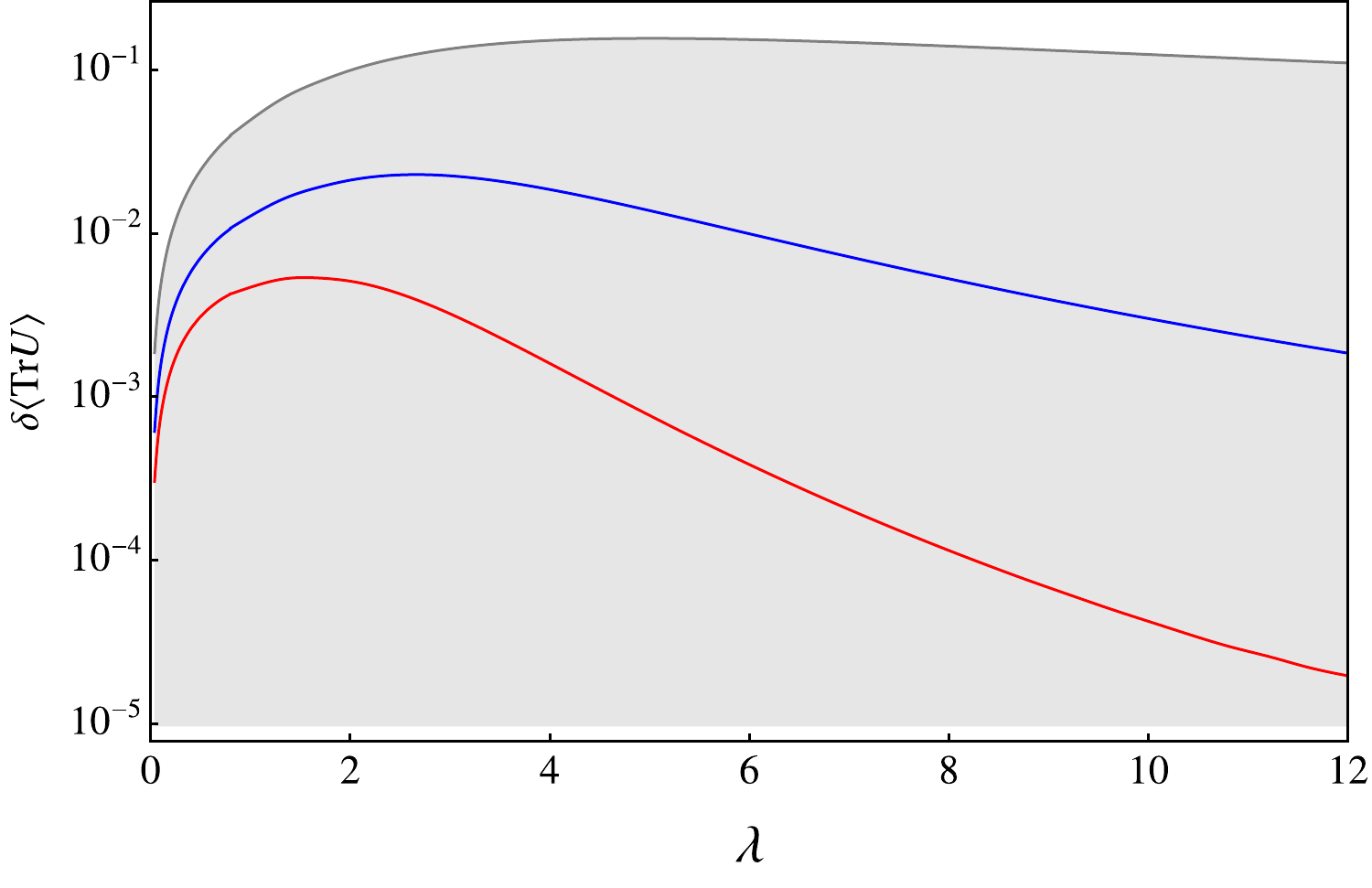}
\end{subfigure}
\caption{The results of our bootstrap calculations in two dimensions for the plaquette average \( u_{P} = \langle \Tr U_P \rangle \) as a function of the coupling \(\lambda = \frac{2N^2}{\beta}\) are depicted in the following manner: Blue lines represent interpolations of the upper and lower margins for the maximum length of loops \( L_{\text{max}} \leq 16 \) i.e. \(\Lambda = 2\) used in the bootstrap procedure for loop equations. Red lines illustrate the same for \( L_{\text{max}} \leq 24 \) or \(\Lambda = 3\). The gray area denotes the allowed values of \( u_{P} \) for \( \Lambda=1\). The black dotted line indicates the exact value, the explicit form of which can be found in Appendix~\ref{app: exact}.
The right plot illustrates the difference between the upper bound and the exact value, demonstrating an exponential convergence behavior.}
\label{fig: 2dsu2}
\end{figure}

Utilizing the hierarchical system, it becomes straightforward to derive relatively tight bounds on observables in two-dimensional lattice gauge theory. Up to $\Lambda=3$, which includes Wilson loops of maximum length 24, we analyze a total of 8335 Wilson loops. These loops are subject to 14591 loop equations\footnote{This indicates a significant level of linear dependency within the space of loop equations.}. After processing, it is found that 7291 of these loops can be linearly represented by the remaining 1044 loops. The positivity condition for this setup involves 18 blocks, with sizes given by:
\begin{equation}
    54, 52, 46, 45, 98, 45, 46, 52, 53, 98, 30, 27, 21, 24, 16, 11, 8, 12
\end{equation}
This configuration presents a complex, yet manageable, bootstrap problem. With the aid of highly optimized software, the entire computation for the $\Lambda=3$ level, which includes 30 pairs of bounds, is completed within one minute\footnote{The solver for SDP we were using for the entire project is \href{https://www.mosek.com/}{MOSEK}.}. The results of the bounds and their convergence are depicted in Figure~\ref{fig: 2dsu2}.

\subsection{Higher dimensional bound and string tension}\label{sec: higherdim}

The entire framework developed for the two-dimensional scenario can, in principle, be extended to three and four dimensions without substantial conceptual challenges. However, practical implementation in higher dimensions poses significantly greater complexity. The complexity of extending our framework to three and four dimensions arises from two primary factors. Firstly, as the spacetime dimension increases, the number of Wilson loops grows exponentially with our truncation level, leading to a significantly larger computational space. Secondly, the loop equations for the three and four-dimensional lattice Yang-Mills theory become considerably more complex. This complexity manifests as a greater number of independent Wilson loops that must be considered once the loop equations are applied. Due to these practical difficulties, specific adaptations were necessary for conducting the \(\Lambda=3\) calculations, where \(L_{\text{max}}=24\). The actual strategy we adopted involved a further truncation of the problem. Specifically, we considered only half of the \(\Lambda=3\) Wilson lines in our calculations. Moreover, in the four-dimensional analysis, we focused exclusively on the reflection positivity aspect. This selective consideration was essential to manage the computational complexity and ensure the feasibility of the calculations within available resources. 

We provide a detailed benchmark of our computational framework for $\Lambda=3$ bootstrap in different spacetime dimensions in Table~\ref{table: technical}\footnote{The bootstrap calculations at the \(\Lambda=2\) level are completed almost instantaneously, regardless of the spacetime dimension under consideration.}. This benchmark highlights the performance metrics such as computational time and memory usage, contextualized by the specific demands of each dimensional analysis\footnote{The time consumption figures are based on executions performed using MOSEK on a machine equipped with two AMD EPYC 7282 processors.}:

- \(\#var\): Represents the total number of Wilson loops considered.
- \(\#LP\): Denotes the number of loop equations incorporated into the analysis.
- \(\#SDPvar\): Indicates the number of free variables remaining after the application of loop equations.
- \(\#Block\) and \(\text{Blocksize}\): These specify the size and number of blocks in the semidefinite programming; for instance, in the four-dimensional case, this involves 40 positive semidefinite matrices, each approximately \(300 \times 300\) in size.
- The final two columns detail the computational resources required for a single optimization, specifically the time and memory consumption, using the MOSEK optimization software to solve the problem.

\begin{table}[t]
\centering
 \begin{tabular}{||c c c c c c c c||} 
 \hline
 Dim & \#var & \#LE  & \#SDPvar & \#Block & Blocksize & Memory & Time \\ [0.5ex] 
 \hline\hline
 2   & 8335 & 14591  & 1044 & 18 & $\sim 50$ & $\sim 100$ MB& $\sim 1$ s \\
 3   & 174387 & 98032   &93561& 38 & $\sim 200$ & 220 GB& $\sim 1$ days \\
 4   & 343851 & 152149 & 211912 & 40 & $\sim 300$ & 1 TB& $\sim 20$ days \\ [1ex] 
 \hline
 \end{tabular}
 \caption{The scale of the \(\Lambda=3\) bootstrap problem we addressed in this project. Here \#var is the number of all the Wilson loops, \#LE is the number of loop equations we considered, \#SDPvar is the number of the free variables after imposing the loop equation. \#Block and Blocksize indicate the size of the semidefinite programming, four example, in four dimensions, it means 40 positive semidefinite matrices with size around $300\times 300$. The last two columns are the time and memory consumption using MOSEK to solve the optimization problem.   }
\label{table: technical}
\end{table}
\begin{figure}[t]
\begin{center}
\includegraphics[scale=0.6]{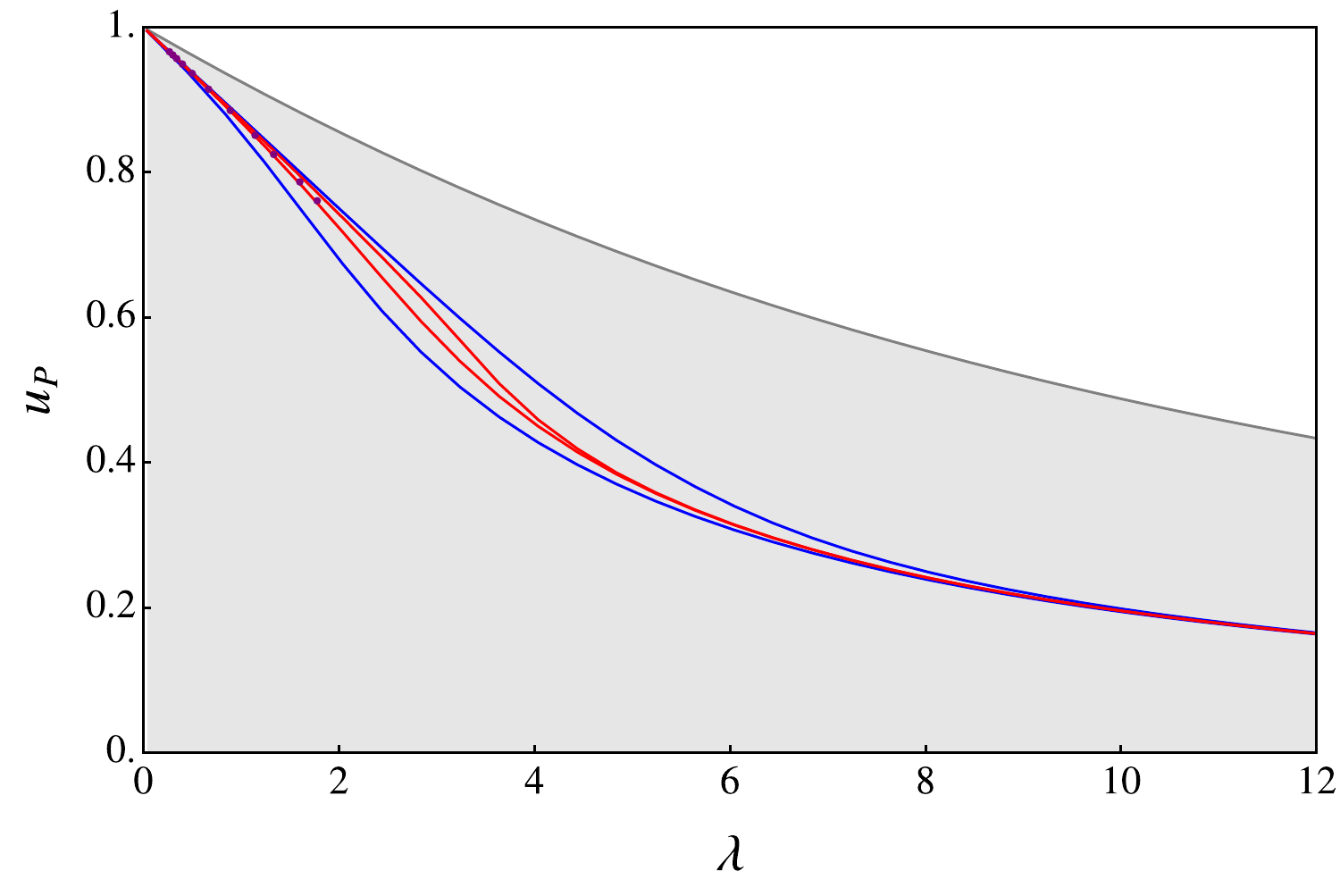}
\end{center}
\caption{The results of our bootstrap calculations in three dimensions for the plaquette average \( u_{P} = \langle \Tr U_P \rangle \) as a function of the coupling \(\lambda = \frac{2N^2}{\beta}\) are depicted in the following manner: Blue lines represent interpolations of the upper and lower margins for the maximum length of loops \( L_{\text{max}} \leq 16 \) i.e. \(\Lambda = 2\) used in the bootstrap procedure for loop equations. Red lines illustrate the same for \( L_{\text{max}} \leq 24 \) or \(\Lambda = 3\). The gray area denotes the allowed values of \( u_{P} \) for \( \Lambda=1\). The purple points are the results of Monte Carlo simulation from~\cite{Athenodorou:2016ebg}.For a more detailed view of the current plot, please refer to Figure~\eqref{fig: intro3}.}  \label{fig: 3dsu2}
\end{figure}

In the three-dimensional analysis presented in Figure~\ref{fig: 3dsu2}, we observe that even in the worst-case scenario, the allowed region for our results remains within around 0.1\% of the exact value. Furthermore, the Monte Carlo data from \cite{Athenodorou:2016ebg} aligns perfectly with our bootstrap bounds, underscoring the robustness and accuracy of our computational approach.

\begin{figure}[t]
\begin{center}
\includegraphics[scale=0.45]{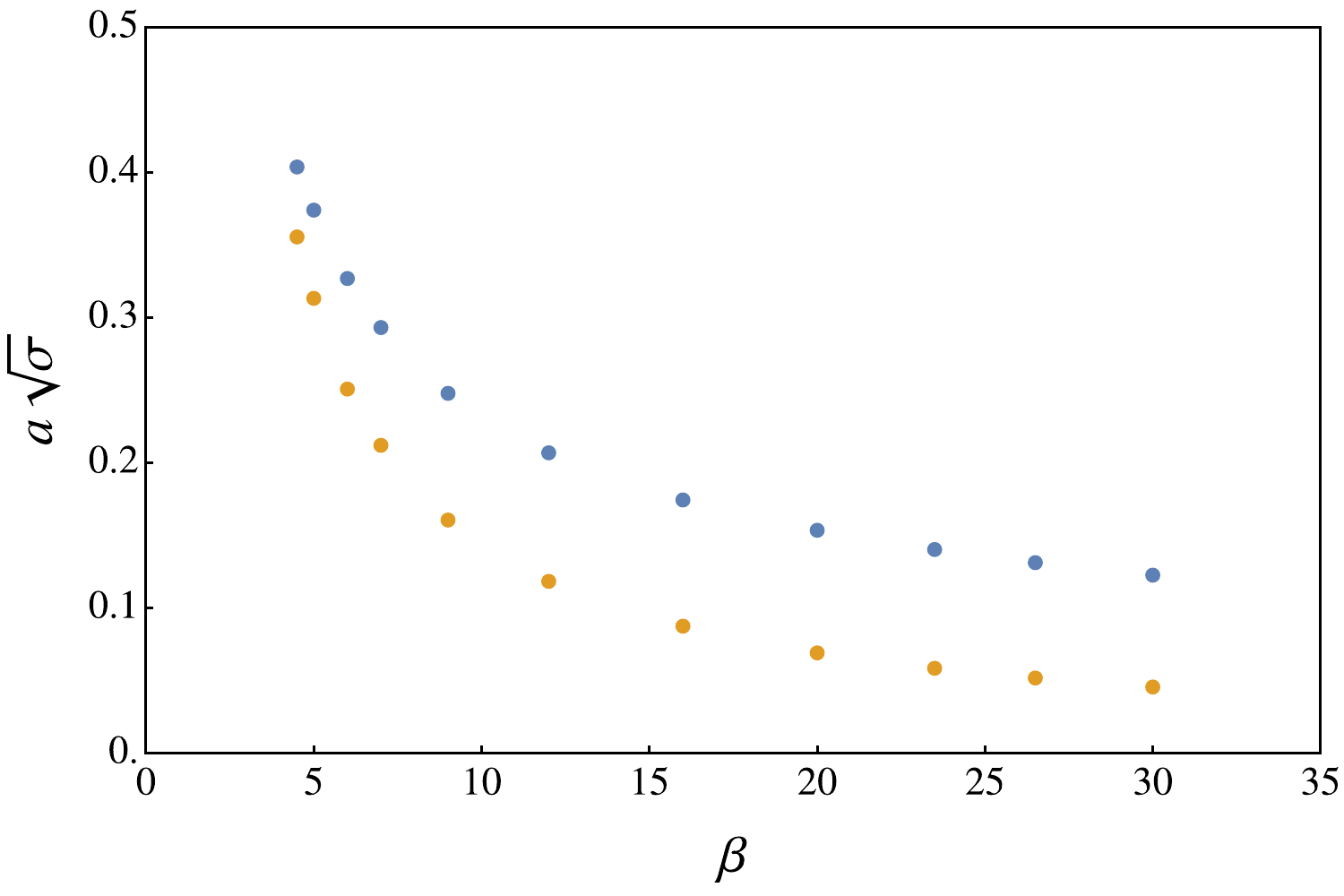}
\end{center}
\caption{The results of the string tension calculations from the \( (2,3) \) rectangular Wilson loop are presented here, with a reminder that \( \lambda = \frac{2N^2}{\beta} \). The brown dots represent Monte Carlo simulation results from \cite{Athenodorou:2016ebg}, whereas the blue dots depict string tension extracted from bootstrap expectations for the loop configurations \( (2,3), (2,2), (3,1), (2,1) \).}  \label{fig: string_tension}
\end{figure}

\begin{figure}[t]
\begin{center}
\includegraphics[scale=0.6]{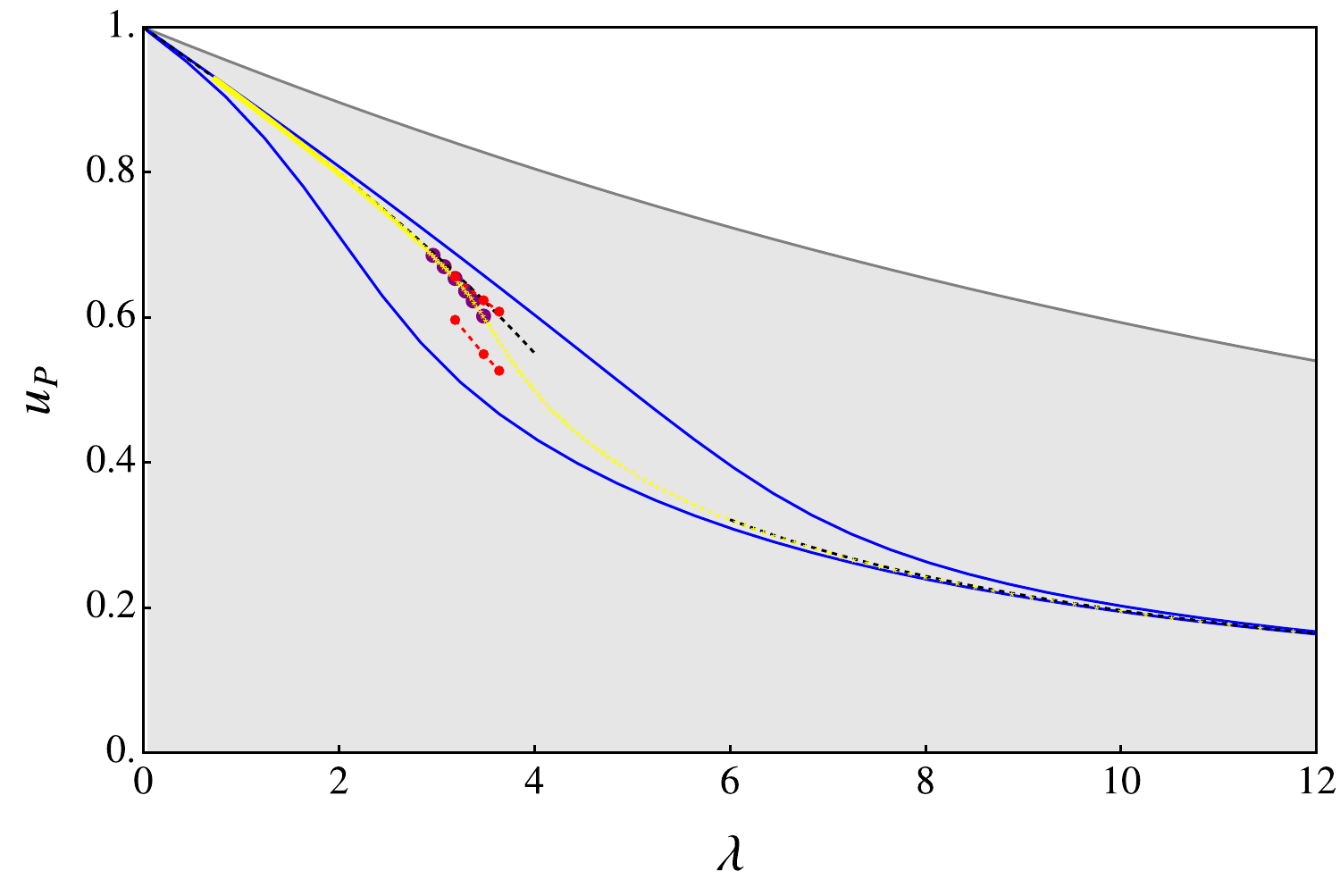}
\end{center}
\caption{The results of our bootstrap calculations in four dimensions for the plaquette average \( u_{P} = \langle \Tr U_P \rangle \) as a function of the coupling \(\lambda = \frac{2N^2}{\beta}\), ($N=2$), in the SU(2) lattice gauge theory are depicted in the following manner: Blue lines represent interpolations of the upper and lower margins for the maximum length of loops \( L_{\text{max}} \leq 16 \) i.e. \(\Lambda = 2\) used in the bootstrap procedure for loop equations. Three pairs of red points illustrate the upper and lower bounds of \( u_P \) for \( L_{\text{max}} \leq 24 \) or \(\Lambda = 3\) at three different values of the coupling \(\lambda=3.19, 3.48, 3.64\). The gray area denotes the allowed values of \( u_{P} \) for \( \Lambda=1\). The blacked-dashed lines in the figure represent the strong and weak coupling expansions, extracted from \cite{Denbleyker:2008ss}. Additionally, the results from Monte Carlo simulations are depicted using purple dots \cite{Athenodorou:2021qvs} and yellow dots \cite{Denbleyker:2008ss}, each set representing data from different sources. For a more detailed view of the current plot, please refer to Figure~\eqref{fig: intro4}.
}  \label{fig: 4dsu2}
\end{figure}

It is commonly accepted that for the Wilson averages \( W[T,L] \) of rectangular Wilson loops of considerable size \( T \times L \), particularly in the regime where \( T \gg L \gg 1 \), the asymptotic behavior is expressed as:
\begin{align}\label{confinement}
    \log W[T,L] = -\sigma TL + \mathcal{O}(1/L)\,.
\end{align}
The first term in this equation is indicative of the confinement "area law," characterized by the string tension \(\sigma(\lambda)\). The string tension \(\sigma(\lambda)\) is routinely extracted from Monte Carlo (MC) data employing the Creutz ratio\cite{Creutz:1980wj}:
\begin{align}
    a^2\sigma = -\log\frac{W(T,L)W(T-1,L-1)}{W(T-1,L)W(T,L-1)}
\end{align}
Although the loops measured in our bootstrap procedure are relatively small compared to those used in various MC simulations for extracting \(\sigma\), it is believed (as discussed in~\cite{Doran:2009pp}) that the regime \eqref{confinement} is attained from very small physical distances, of the order of 0.5 fm. Consequently, we attempt here to extract, at least qualitatively, \(\sigma(\lambda)\) from our bootstrap data for loops of size \( (T, L) = (3,2) a \). In our analysis, we select the optimal solution corresponding to the lower bound of the one-plaquette Wilson loop and use it to extract the expectation values for the loop configurations \( (2,3), (2,2), (3,1), (2,1) \). While the choice to focus on the lower bound might appear arbitrary, it is important to note that the expectation values exhibit minimal variation when the optimal solution for the upper bound is used instead. This stability between the lower and upper bounds is corroborated by the similar allowed margins for larger Wilson loops, relative to those for the one-plaquette Wilson loop, as demonstrated in Figure~\ref{fig: 3dsu2}. This consistency across different bounds and loop sizes not only validates our methodological approach but also reinforces the reliability of our bootstrap predictions. This approach ensures that our results are robust, minimizing the impact of the specific choice of lower or upper bounds on the final outcomes. 

The comparison of our results with those from Monte Carlo simulations \cite{Athenodorou:2016ebg} is depicted in Figure~\ref{fig: string_tension}. A notable observation is that for larger values of \(\beta\), where the physical scale of lattice spacing \(a\) is relatively small, our agreement with the Monte Carlo data is not perfect. Conversely, for smaller \(\beta\), which corresponds to a larger lattice spacing \(a\), there is a decent match with the Monte Carlo results. This outcome aligns with expectations, as precise estimations of the string tension from our bootstrap method are anticipated only when the product \((3,2) \times a\) falls within the asymptotic region described by \eqref{confinement}. This comparison not only illustrates how bootstrap data correlates with established Monte Carlo simulations but also provides qualitative validation of our bootstrap method for extracting the string tension, even with relatively small loop sizes. 

The bootstrap results for the four-dimensional lattice Yang-Mills theory are depicted in Figure~\ref{fig: 4dsu2}. We compare these results with perturbation theory \cite{Denbleyker:2008ss} and Monte Carlo data from various sources \cite{Athenodorou:2021qvs, Denbleyker:2008ss}. The comparison shows that both the Monte Carlo and perturbation theory results fall well within the region allowed by our bootstrap method, affirming the validity of our approach. 

Unlike the scenarios in two and three dimensions, for \(\Lambda=3\) or \(L_{\text{max}}=24\), we derived only three pairs of bounds in the four-dimensional bootstrap. This limitation reflects the maximum computational capacity achievable with our current software and hardware setups. Moreover, we observed a slight lack of numerical precision in the lower bounds, suggesting that merely increasing computational resources might not significantly enhance the results. This observation underscores the need for possibly refining our computational strategies or methodologies to overcome these challenges in higher-dimensional settings.

\section{Discussion and Prospects}

Although our current exploration is limited to SU(2) lattice gauge theory, we believe that our approach applies to a more general framework encompassing SU(N) and U(N) gauge theories, both theoretically and practically. The theoretical applicability is evident and further substantiated by our explorations of SU(3) and U(2) one-matrix models. Practically, the feasibility of broader application is supported by a simple yet significant observation: within a fixed truncation level in our hierarchy, the single trace Wilson loop consistently dominates. The most advanced result achieved in our project so far is at truncation level \(\Lambda=6\), leading us to anticipate the possibility of establishing a non-trivial bound for SU(6) or U(5). Here, the lowest non-trivial five-trace loop equation corresponds to level 6 in our hierarchy. For higher values of \(N\), while valid bounds can still be obtained, their dependence on \(N\) tends to be trivial, as even the lowest order trace identity does not manifest below level \(\Lambda=6\).

Our current project has not overcome the main bottleneck facing the state of the art in lattice bootstrap, as it was already manifest in previous studies \cite{Anderson:2016rcw, Kazakov:2022xuh, Cho:2022lcj}: our analysis remains confined to a local region of the infinite lattice, despite notable technical achievements. To significantly advance beyond this limitation, we foresee the potential for implementing a coarse-graining procedure analogous to the Density Matrix Renormalization Group (DMRG) used in quantum spin chain studies \cite{White:1992zz}. This approach has recently shown substantial progress in related fields \cite{Kull:2022wof, DMRGBootWIP}. Additionally, a more detailed understanding of the close-to-null operators in lattice gauge theory could lead to the development of specific functionals based on these operators. Recent advances in conformal bootstrap suggest that such strategies could considerably enhance analytic efficiency \cite{Ghosh:2023onl}.

It would be exciting to extend our framework from the statistical ensemble to quantum systems involving finite \(N\) matrix models. Particularly in scenarios like the finite SU(2) matrix quantum mechanics, where even the ground state remains inadequately understood \cite{Hoppe:2023qem}, our approach could provide significant insights. Given the rigor of the bounds established by our method, it could offer valuable complementary information to the numerical studies of matrix quantum mechanics, especially those related to gravity  duals of such models \cite{Komatsu:2024vnb}. Such an extension could enhance our understanding of quantum behavior in matrix models and potentially contribute to a broader field of quantum gravity research.

In addition, we aim to elucidate the effectiveness of reflection positivity, which has remained somewhat enigmatic. Previous studies, such as \cite{Kazakov:2021lel} on the matrix model and \cite{Cho:2023ulr} on the lattice Ising model, have clarified the role of positivity from Hermitian conjugation; it is closely associated with the absence of a sign problem in the model under consideration. It is generally believed that merely imposing Hermitian conjugation positivity ensures the convergence of the bootstrap procedure. However, reflection positivity extends beyond this by guaranteeing the possibility of defining a physical Hilbert space on the spatial reflection plane, which suggests the potential for a Hamiltonian formulation of the theory in the infrared regime. This observation also hints that the bootstrap method on the lattice could potentially transcend the limitations imposed by the sign-problem, which is prevalent in the Monte Carlo simulation methods. This suggests a promising avenue for advancing non-perturbative studies in lattice theories without the computational burdens typically associated with sign issues.

The convergence of the bootstrap based on the reflection positivity is also non-conclusive. In a previous study of the large \(N\) Lattice Yang-Mills theory \cite{Kazakov:2022xuh}, it was observed that imposing reflection positivity as the sole type of positivity condition did not yield any non-trivial upper bounds for \(\langle \Tr U \rangle\). However, in our current project and in the previous study of the lattice Ising model across two and three spacetime dimensions \cite{Cho:2022lcj}, we have observed that reflection positivity exhibits a markedly greater influence than Hermitian conjugation positivity. This observation suggests that reflection positivity may play a more pivotal role in shaping the dynamics and theoretical outcomes of such models.

The fact that SU(2) loop equations are linear and closed on the single-loop space (and similarly, the linear SU(3) loop equations are closed on the two-loop space) can have, in principle an important impact on the understanding of the structure of non-abelian gauge theories in general. It would be important to understand what are the redundancies for these equations in the loop space and what kind of boundary conditions (or maybe asymptotics on large loops) we should impose to single out the relevant physical solutions. These insights could also significantly improve the current algorithms and may enable an efficient application of neural network methods, to try to extrapolate our results to longer Wilson loops.   

An important problem is to include the dynamical quarks into the bootstrap. In principle, the multi-loop equations allow the inclusion of internal quark loops and then sum up their shapes to render the physical results. Moreover, for SU(2) case all such multiloop configurations are in principal generated by single loop Wilson averages. A similar, though more involved statement is true for the two-loop space of SU(3) gauge theory. However, the practical bootstrap computations in this direction may be very involved.

It is also important to note that our formalism allows us to approach the baryonic loop averages, as defined and formulated in~\cite{Kazakov:1981zs}.

\section*{Acknowledgement}
We are grateful for the discussions with Yiming Chen, Minjae Cho, Barak Gabai, Nikolay Gromov, Martin Kruczenski, Henry Lin, Raghu Mahajan, Alexandre Migdal, Colin Oscar Nancarrow, Wenliang Li, Zhijin Li, Ying-Hsuan Lin, Victor A. Rodriguez, Joshua Sandor, Stephen Shenker, Evgeny Sobko, Douglas Stanford, Yuan Xin and Xi Yin. 
V.K. thanks the Perimeter Institute for Theoretical Physics and the Simons Center for Geometry and Physics, where a part of this work has been done.
Z.Z. would like to express his sincere gratitude to Pedro Vieira, who provided invaluable support and guidance during his stay at ICTP-SAIFR, S\~{a}o Paulo, where the final stage of the project is finished.  Z.Z is supported by Simons Foundation grant \#994308 for the Simons Collaboration on Confinement and QCD Strings. 

\section*{Appendices}

\appendix

\section{Exact results for the unitary one-matrix model}\label{app: exact}
In this section, we present some exact results of the one-matrix model discussed in the main text. The partition function \(Z\) is defined as:
\begin{equation}
    Z=\int \mathrm{d}U\, e^{\frac{N}{\lambda}(\text{tr} U^\dagger +\text{tr} U)}
\end{equation}
where the Haar measure is given by:
\eqsplit[]{
\mathrm{d}U_{U(N)}&=\frac{1}{N!}\prod_{1\leq j<k\leq N}\left((e^{i\theta_j}-e^{i\theta_k})(e^{-i\theta_j}-e^{-i\theta_k})\right)\prod_{j=1}^N \frac{\mathrm{d}\theta_j}{2\pi},\\
\mathrm{d}U_{SU(N)}&=\frac{1}{N!}\sum_q \delta\left(\sum_{j=1}^N \theta_j-2q\pi\right)\prod_{1\leq j<k\leq N}\left((e^{i\theta_j}-e^{i\theta_k})(e^{-i\theta_j}-e^{-i\theta_k})\right)\prod_{j=1}^N \frac{\mathrm{d}\theta_j}{2\pi}.
}
The integration is performed from \(0\) to \(2\pi\) for each \(\theta_i\).

The exact values for the partition function are known \cite{Jha:2020aex}:
\eqsplit[]{
Z_{U(N)}(N, \lambda)&=\text{Det}\left[I_{j-k}\left(\frac{2N}{\lambda}\right)\right]_{j, k=1,\ldots,N},\\
Z_{SU(N)}(N, \lambda)&=\sum_{p=-\infty}^{\infty} \text{Det}\left[I_{j-k+p}\left(\frac{2N}{\lambda}\right)\right]_{j, k=1,\ldots,N}.
}
For the special case of \(SU(2)\), the infinite sum can be explicitly calculated:
\begin{equation}
    Z_{SU(2)}(2, \lambda)=\frac{1}{4} \lambda I_1\left(\frac{8}{\lambda}\right),
\end{equation}
where \(I_n\) represents the modified Bessel function of the first kind, as defined, for example, in the BesselI function in Mathematica.

The single-trace moments can be derived by taking derivatives of the partition function. It can be directly verified that both \(\langle \text{Tr} U\rangle_{SU(N)}\) and \(\langle \text{Tr} U\rangle_{U(N)}\) converge to the known analytic result in the 't Hooft limit \cite{Gross:1980he, Wadia:2012fr}:
\begin{equation}
    \langle \text{Tr} U\rangle_{SU(\infty)}= \begin{cases}1-\frac{\lambda}{4}, & \text{for } 0\leq \lambda \leq 2 \\ \frac{1}{\lambda}, & \text{for } \lambda \geq 2\end{cases}.
\end{equation}
For various values of \(N\), \(\langle \text{Tr} U\rangle_{SU(N)}\) monotonically decreases, whereas \(\langle \text{Tr} U\rangle_{U(N)}\) monotonically increases.
\section{Loop equations for SU(N)/U(N) one-matrix model}\label{app: generalloop}
In this appendix, we provide the derivation of the multi-trace loop equations for the following model:
\begin{equation}
    Z=\int \mathrm{d}U\, e^{\frac{N}{\lambda}(\text{tr} U^\dagger +\text{tr} U)}
\end{equation}
This framework is extended to accommodate more general cases of SU(N)/U(N) matrix model.

The derivation of loop equations for this model is crucial for understanding the dynamics within our bootstrap approach and can be represented as follows:
\begin{equation}
    \int  \mathcal{D} U\, \delta_\epsilon\left(  U^n_{ab} \Tr U^{m_1}...\Tr U^{m_{N-2}} \e^{\frac{N}{\lambda}( \tr U^\dagger +\tr U)}\right)=0
\end{equation}
The loop equations are derived using the concept of small left variations on the group, denoted as \(\delta_\epsilon\). This variation is defined as follows:
\begin{equation}
    \delta_\epsilon (U_{ab})=U_{ab}+i\epsilon_{ac}U_{cb},\qquad\delta_\epsilon (U^\dagger_{ab})=U^\dagger_{ab}-iU^\dagger_{ac}\epsilon_{cb}
\end{equation}

Here, \(\epsilon\) represents an arbitrary traceless Hermitian matrix in the context of SU(N) gauge theory, and an arbitrary Hermitian matrix for U(N) scenarios. This distinction ensures that the variations are appropriately aligned with the structural properties of each group:
\eqsplit[]{
&\epsilon_{cd} \langle\sum_{i=1}^{N-2} \frac{m_i}{N} U^n_{ab} U^{m_i}_{dc}\Tr U^{m_1}...\Tr U^{m_{i-1}} \Tr U^{m_{i+1}} ...\Tr U^{m_{N-2}} \\
&+\sum_{i=0}^{n-1} U^i_{ac} U^{n-i}_{db}\Tr U^{m_1}...\Tr U^{m_{N-2}}+\frac{N}{\lambda} U^n_{ab} \Tr U^{m_1}...\Tr U^{m_{N-2}} (U_{dc}-U^\dagger_{dc})\rangle=0
}

We observe that if \(\epsilon_{cd} X_{dc} = 0\) for any \(\epsilon\), then it implies that \(X_{dc} = \frac{s_G}{N} \delta_{dc} X_{ff}\). Here, \(s_G\) is a scalar factor specific to the group in question and is defined in \eqref{eq: sg} in the main text. This relationship can be expressed as follows:
\eqsplit[]{
&
\langle\sum_{i=1}^{N-2} \frac{m_i}{N} U^n_{ab} U^{m_i}_{dc}\Tr U^{m_1}...\Tr U^{m_{i-1}} \Tr U^{m_{i+1}} ...\Tr U^{m_{N-2}} \\
&+\sum_{i=0}^{n-1} U^i_{ac} U^{n-i}_{db}\Tr U^{m_1}...\Tr U^{m_{N-2}}+\frac{N}{\lambda} U^n_{ab} \Tr U^{m_1}...\Tr U^{m_{N-2}} (U_{dc}-U^\dagger_{dc})\rangle\\
&=s_G\delta_{dc}\langle \frac{1}{N}n U^{n}_{ab}\Tr U^{m_1}...\Tr U^{m_{N-2}}+\frac{N}{\lambda}  U^n_{ab} \Tr U^{m_1}...\Tr U^{m_{N-2}}(\Tr U-\Tr U^\dagger)\\
& +(\sum_{i=1}^{N-2} m_i) U^n_{ab} \Tr U^{m_1}...\Tr U^{m_{N-2}}\rangle
}
Contracting both side by $\frac{1}{N^2} \delta_{ac}\delta_{bd}$, we get~(for $n>0$):
\eqsplit[]{
&
\langle\sum_{i=1}^{N-2} \frac{m_i}{N^2} \Tr U^{n+m_i} \Tr U^{m_1}...\Tr U^{m_{i-1}} \Tr U^{m_{i+1}} ...\Tr U^{m_{N-2}} \\
&+\sum_{i=0}^{n-1} \Tr U^i \Tr U^{n-i} \Tr U^{m_1}...\Tr U^{m_{N-2}}+\frac{1}{\lambda}  \Tr U^{m_1}...\Tr U^{m_{N-2}} (\Tr U^{n+1}-\Tr U^{n-1})\rangle\\
&=\frac{s_G}{N^2}\langle n \Tr U^{n}\Tr U^{m_1}...\Tr U^{m_{N-2}}+\frac{N^2}{\lambda}  \Tr U^n \Tr U^{m_1}...\Tr U^{m_{N-2}}(\Tr U-\Tr U^\dagger)\\
& +(\sum_{i=1}^{N-2} m_i) \Tr U^n \Tr U^{m_1}...\Tr U^{m_{N-2}}\rangle
}

\section{More bootstrap of the one-matrix model}\label{app: moreboot}

\begin{figure}
\centering
\begin{subfigure}{.5\textwidth}
  \centering
  \includegraphics[width=\linewidth]{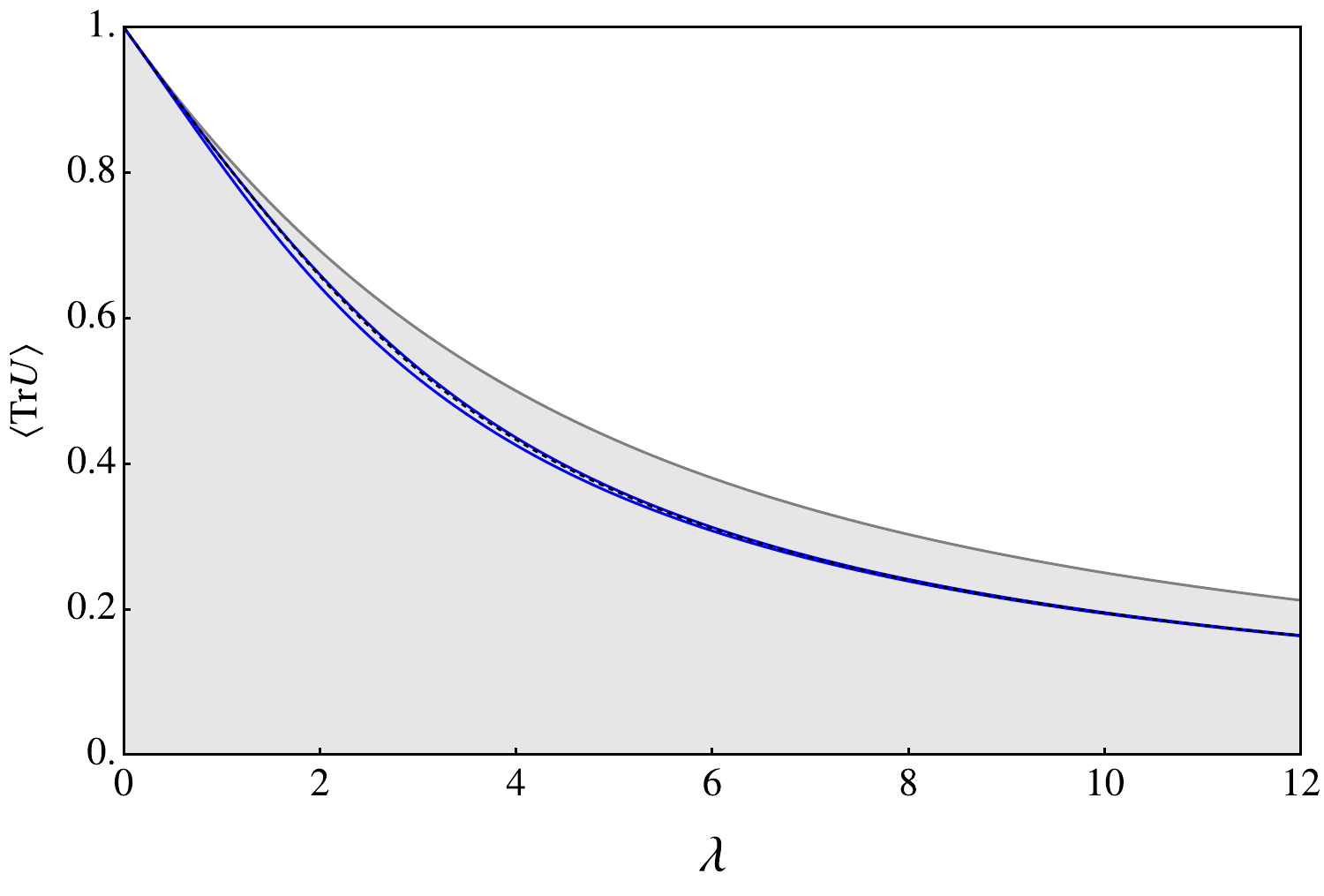}
\end{subfigure}%
\begin{subfigure}{.5\textwidth}
  \centering
  \includegraphics[width=\linewidth]{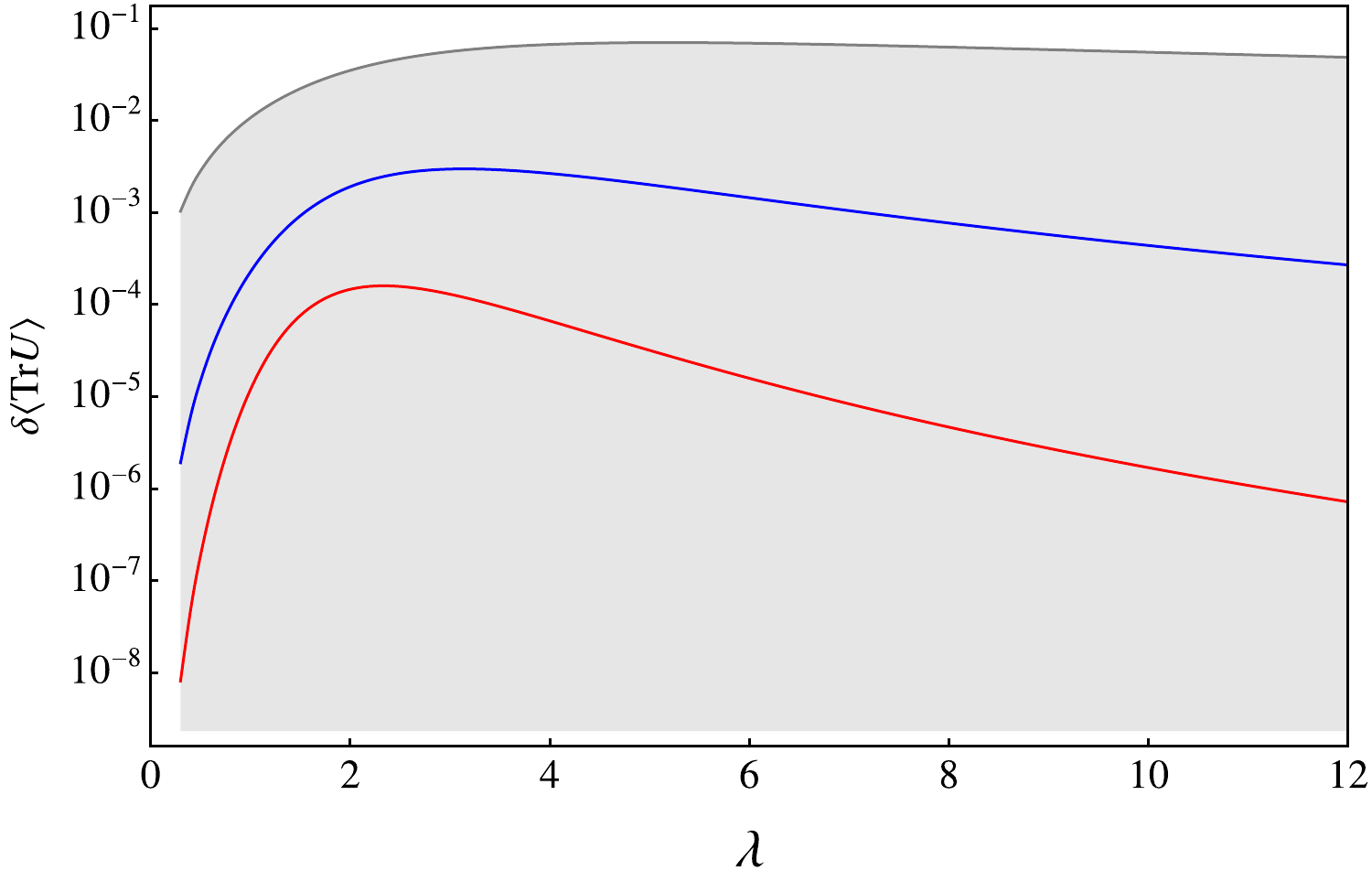}
\end{subfigure}
\caption{SU(2) bounds and convergences.}
\label{fig: toysu2}
\end{figure}

\begin{figure}
\centering
\begin{subfigure}{.5\textwidth}
  \centering
  \includegraphics[width=\linewidth]{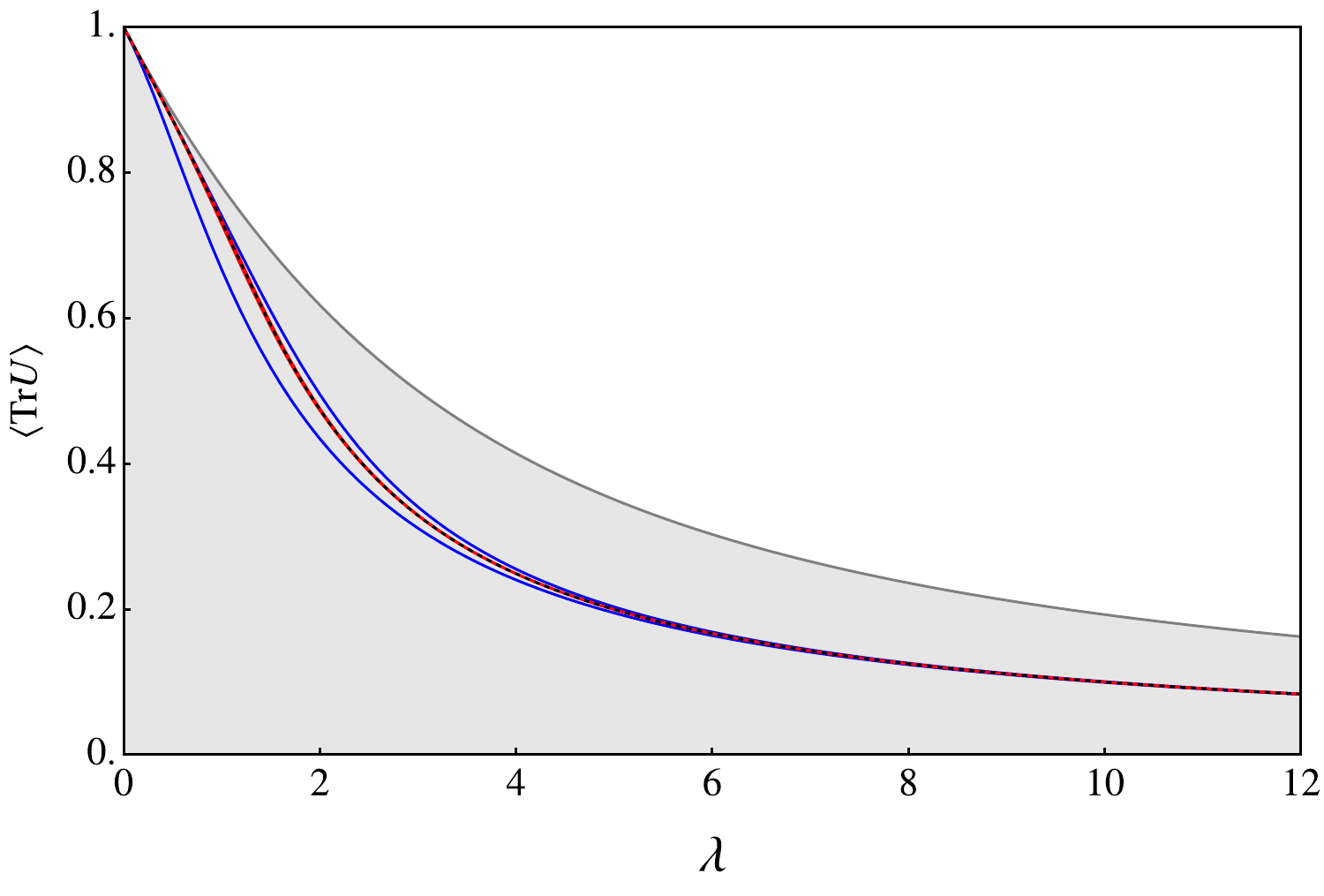}
\end{subfigure}%
\begin{subfigure}{.5\textwidth}
  \centering
  \includegraphics[width=\linewidth]{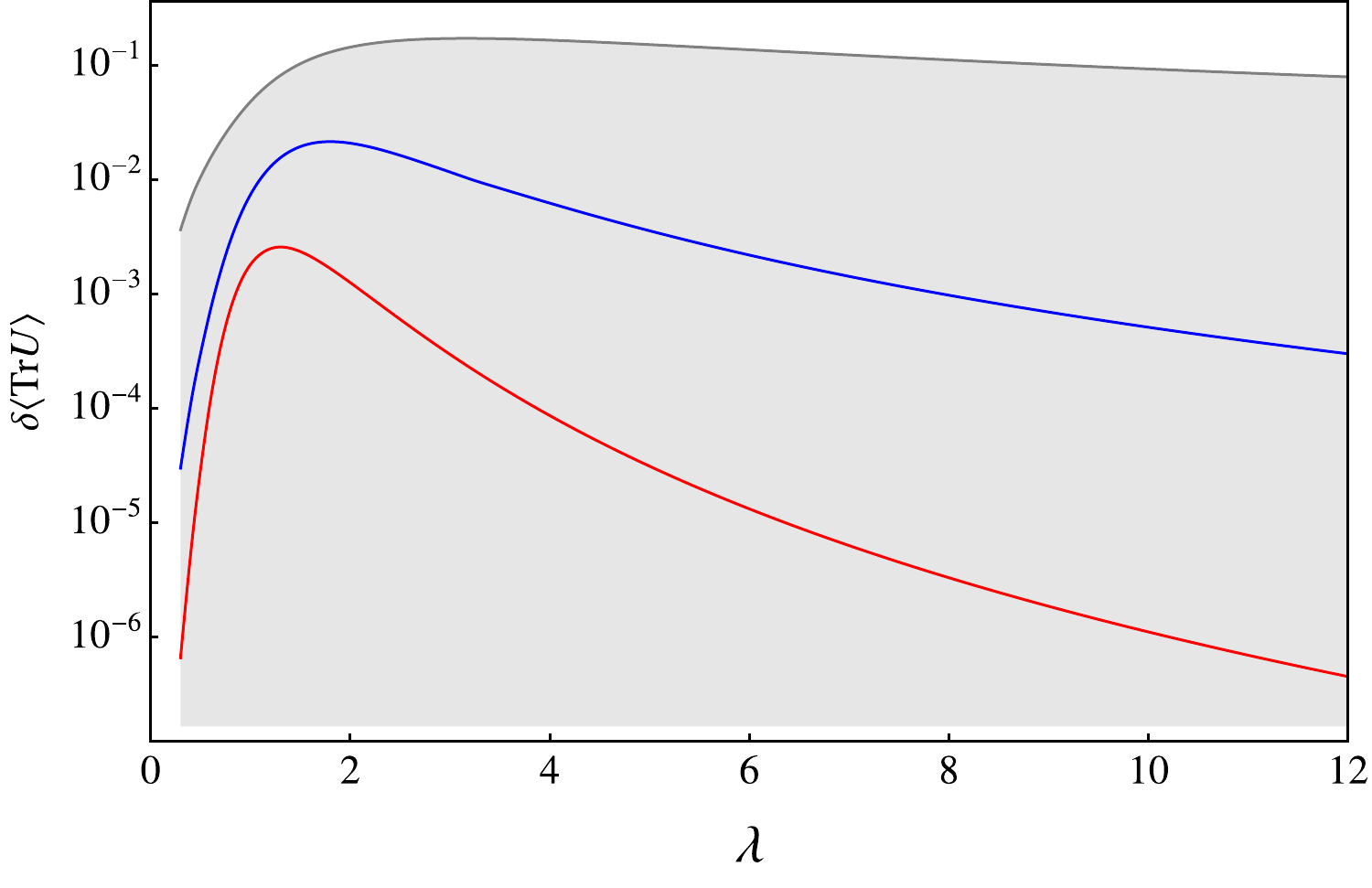}
\end{subfigure}
\caption{U(2) bounds and convergences.}
\label{fig: toyu2}
\end{figure}

\begin{figure}
\centering
\begin{subfigure}{.5\textwidth}
  \centering
  \includegraphics[width=\linewidth]{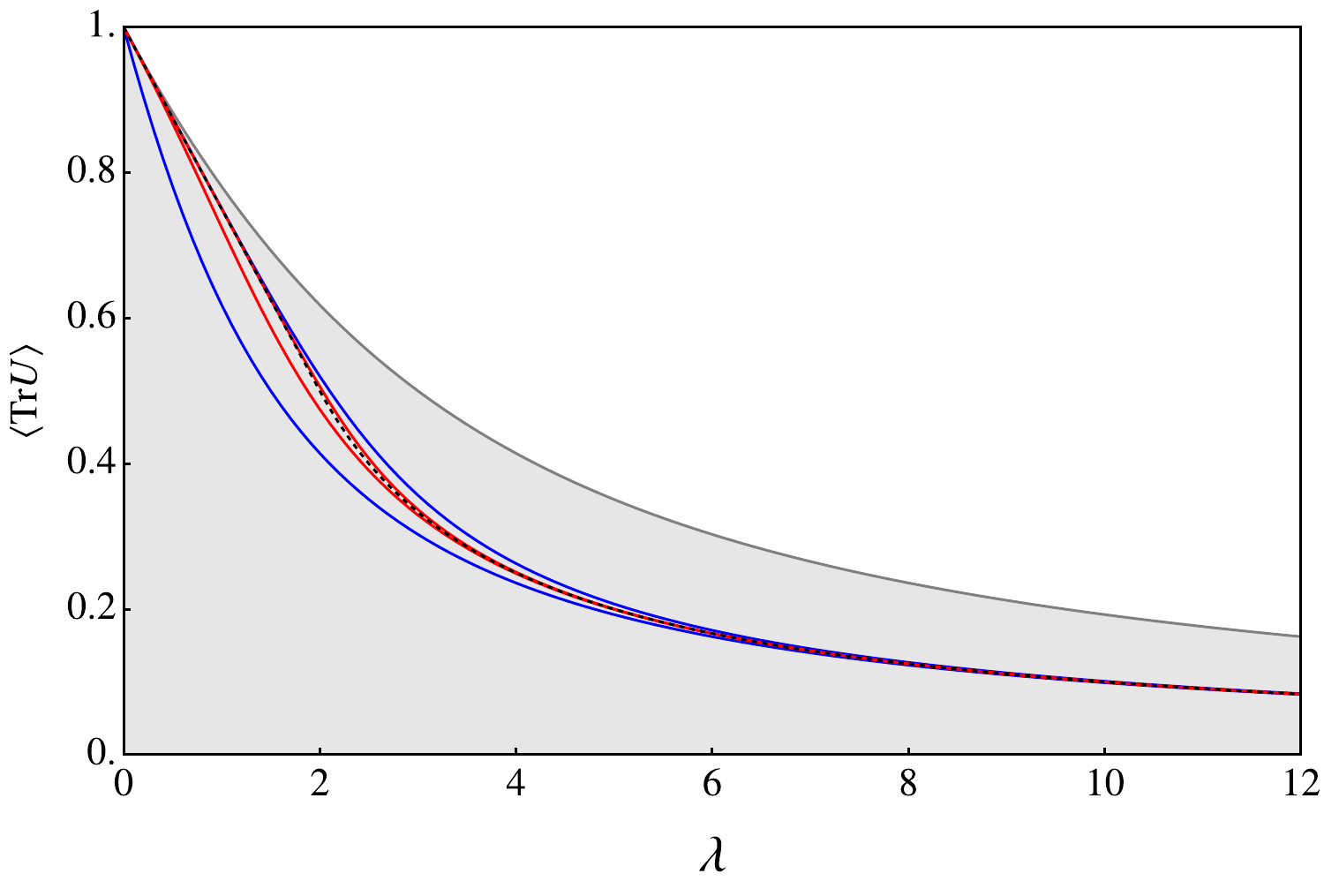}
\end{subfigure}%
\begin{subfigure}{.5\textwidth}
  \centering
  \includegraphics[width=\linewidth]{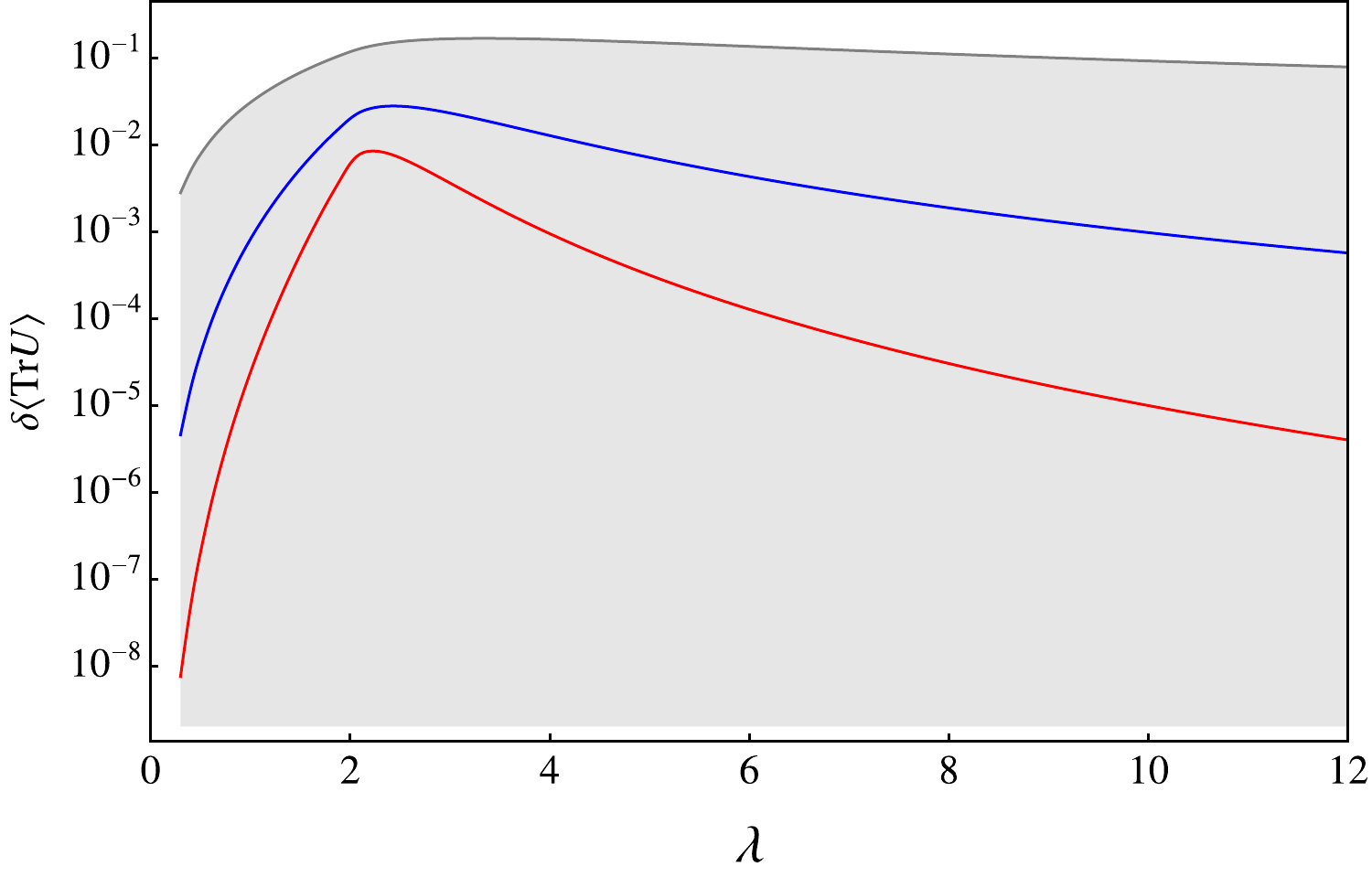}
\end{subfigure}
\caption{SU($\infty$) bounds and convergences.}
\label{fig: toysuinf}
\end{figure}

In this appendix, we present additional results from the bootstrap analysis of the one-matrix model \eqref{eq: onematrixsun}. The corresponding figures, Figure~\ref{fig: toysu2}, Figure~\ref{fig: toyu2}, and Figure~\ref{fig: toysuinf}, represent the SU(2), U(2), and U($\infty$) counterparts of Figure~\ref{fig: toysu3} discussed in the main text. We offer the following remarks on these results:
\begin{itemize}
    \item The structure of the loop equations reveals that all moments are functions of $\langle \Tr U \rangle$ for both SU(2) and U($\infty$)\footnote{However, it is important to note that these are polynomials of $\langle \Tr U \rangle$ in the case of U($\infty$).}. In contrast, the U(2) model requires two variables to determine the rest: $\langle \Tr U \rangle$ and $\langle \Tr U \Tr U^\dagger \rangle$, which closely resembles the SU(3) scenario.
    \item The convergence behavior of the U(2) bootstrap closely parallels that of the SU(3) model discussed in the main text, exhibiting similar rates. However, the SU(2) bootstrap converges significantly faster. Notably, the U($\infty$) bootstrap shows intriguing characteristics: the upper and lower bounds converge at distinctly different rates, particularly in the weak coupling phase where $\lambda \leq 2$. Additionally, there is a pronounced spike at the phase transition point $\lambda = 2$.
\end{itemize}
\newpage
\bibliography{Finite_N_bootstrap}
\bibliographystyle{jhep}



\end{document}